\documentclass[conference]{IEEEtran}
\IEEEoverridecommandlockouts

\usepackage{cite}
\usepackage{amsmath,amssymb,amsfonts}
\usepackage{algorithmic}
\usepackage{graphicx}
\usepackage{textcomp}
\usepackage{xcolor}
\usepackage[hyphens]{url}

\usepackage{amsmath}

\usepackage{amsfonts}
\usepackage{graphicx}
\usepackage{textcomp}
\usepackage{fancyhdr}
\usepackage{hyperref}
\usepackage{xspace}
\usepackage{tabularx}
\usepackage{booktabs}  
\usepackage{makecell}  
\usepackage{libertine}
\usepackage{pifont}
\usepackage{multirow}
\usepackage{soul}
\usepackage{thmbox}

\usepackage{etoolbox}

\usepackage{tcolorbox}

\usepackage{mdframed}

\newcommand{\halfcheckmark}{\ding{52}\rotatebox[origin=c]{-9.2}{\kern-0.7em\ding{55}}}
\newcommand{\myparagraph}[1]{\vspace {3pt}\noindent\textbf{\emph{#1}}}
\newcommand{\REVISE}[1]{#1}

\newcommand{\sys}{\textsc{CHIME}\xspace}
\newcommand{\syscolor}{\textsc{CHIME}\ }

\definecolor{L3orange}{RGB}{255,179,102}
\definecolor{L3cyan}{RGB}{183,210,254}

\def\BibTeX{{\rm B\kern-.05em{\sc i\kern-.025em b}\kern-.08em
    T\kern-.1667em\lower.7ex\hbox{E}\kern-.125emX}}

\begin{document}

\pdfpagewidth=8.5in
\pdfpageheight=11in

\newcommand{\iscasubmissionnumber}{755}

\pagenumbering{arabic}

\title{\sys: A Case for Efficient Long-Context Attention-FC Disaggregated Inference with DIMM-PIM}
\author{
  \IEEEauthorblockN{
    Qingyuan~Liu\IEEEauthorrefmark{1},
    Liyan~Chen\IEEEauthorrefmark{1},
    Haocheng~Wang,
    Yanning~Yang,
    Dong~Du,\\
    Zhigang~Mao,
    Naifeng~Jing,
    Yubin~Xia,
    Haibo~Chen
  }
  \IEEEauthorblockA{
    Shanghai Jiao Tong University, Shanghai, China
  }
}

\maketitle

\renewcommand{\thefootnote}{\fnsymbol{footnote}}
\footnotetext[1]{Co-first author.}
\renewcommand{\thefootnote}{\arabic{footnote}}

\thispagestyle{plain}
\pagestyle{plain}

\begin{abstract}
Attention-FC Disaggregated (AFD) LLM inference systems offload memory-bound Attention operations to memory-rich accelerators (e.g., CPUs, HBM-PIM) while retaining compute-bound Fully-Connected (FC) operations on GPUs. 
In this paper, we first design a Disaggregated Roofline Model (DRM) to characterize AFD performance, revealing that \emph{system throughput is constrained by the accelerator's limiting factor: either memory bandwidth or capacity}.
We observe that prior AFD systems often overlook these constraints and fail to balance them, leading to resource underutilization or constrained throughput.
Therefore, we propose \sys, the first AFD system integrating \textit{DIMM-PIM}, which is a case of the new accelerator that strikes the balance with scalable capacity and bandwidth.
To address the synchronization challenges inherent to the distributed cooperating DRAM chips in DIMM-PIM, \sys employs \emph{bubble-free pipelining} and \emph{hybrid-grained re-layout} for efficient attention computation.
Furthermore, it maximizes cross-device resource utilization via \emph{rankset-granular communication-computation overlapping} and \emph{alignment-predicting scheduling}. 
Evaluations show \sys achieves up to 5.15$\times$ speedup over state-of-the-art HBM-PIM solutions.

\end{abstract}

\section{Introduction}
\label{sec:intro}

\emph{``A plant's growth is limited by the nutrient in shortest supply, regardless of how abundant other nutrients may be.''}
\begin{flushright}
  ---\emph{Liebig's Law}~\cite{liebigslaw}, Justus von Liebig (1803-1873).
\end{flushright}

The demand for Large Language Models (LLMs) to process and generate longer sequences 
is rapidly increasing, driven by applications like long reasoning and complex code generation~\cite{deepseekai2025deepseekr1,plaat2024reasoning,huang2023reasoning,zhou2024survey,zhong2024evaluation,grattafiori2024llama3}.
In these long-context scenarios, the decoding process becomes a significant performance bottleneck.
Moreover, the increasingly longer KV cache imposes growing memory capacity and bandwidth demands on the inference system.

To tackle this memory challenge, Attention-FC Disaggregation (AFD) has emerged as a promising solution
~\cite{step3blog,step3system,heo2024neupims,park2024attacc,he2025papi,jiang2024neo,zhu2025megascaleinfer}.
AFD systems partition the LLM inference workload: the memory-intensive Attention operations and KV cache are
offloaded to a specialized memory-rich device (hereafter referred to as the \textit{``accelerator''}), while the compute-intensive
Fully-Connected (FC) operations remain on the GPU or NPU.
For example, leveraging HBM-based Processing-in-Memory (HBM-PIM) can dramatically
reduce the latency of Attention~\cite{heo2024neupims,park2024attacc,he2025papi} with significantly increased bandwidth.
In conclusion, AFD systems show the potential for higher throughput.

However, the promise of AFD architecture is not always straightforward to realize.
Our investigation reveals a critical yet counter-intuitive phenomenon: 
\textbf{simply enhancing the accelerator does not always guarantee better system performance.}
In an evaluation on GPT-175B with OpenR1~\cite{open-r1}, we found that equipping DGX-A100's HBMs with PIM from 1$\times$ to 16$\times$ higher bandwidth only leads to \textless 1\% throughput improvement.
This puzzling outcome cannot be explained by existing performance models, such as the classic roofline model~\cite{roofline},
which analyzes devices in isolation and fails to capture the complex interplay between disaggregated components in an AFD system.

To address this, we design the \textbf{Disaggregated Roofline Model (DRM)}(\textsection\ref{s:demys}), 
the first general performance model to analyze how AFD systems perform in various scenarios.
DRM holistically characterizes the performance of both the accelerator and the GPU, crucially modeling their interdependencies.
Using DRM, we conclude a principle guiding the design for AFD systems:
the system's throughput is limited by the scarcest resource, be it the accelerator's memory bandwidth or capacity.
Crucially, improvements to any non-bottleneck resource will result in diminishing or even negligible returns.
It explains the phenomenon described in the previous paragraph, and motivates the design of a new accelerator that achieves a more balanced tradeoff among capacity, bandwidth, and scalability.

In this paper, we show that PIM integrated with \textit{DIMM}, i.e., equipping host memory modules with bank-level processing units, constitutes a case of a new accelerator that strikes this balance.
Therefore, we propose AFD systems that leverage \textit{DIMM-PIM}.
Compared to prior HBM/GDDR-PIM solutions~\cite{heo2024neupims,park2024attacc,he2025papi,gu2025cent}, an AFD system with DIMM-PIM gains three key advantages.
First, it directly alleviates the capacity bottleneck of HBM-PIM while providing significantly higher bandwidth than standard host memory, enabling higher overall throughput.
Second, it inherits the superior scalability and configurability of the standard DIMM interface, making the system adaptable and future-proof for diverse memory demands.
Third, as analyzed in \textsection\ref{sec:motiv-dimm-pim-proposal}, it offers higher economic efficiency, which could be cheaper than HBM/GDDR-PIM and suffer less memory wastage.

However, effectively leveraging DIMM-PIM in AFD systems is non-trivial. 
A naive integration of DIMM-PIM introduces additional synchronization overheads both within the PIM hardware and across the inference system, including transfer, layout, and progress synchronization, which can severely stall inference.
Specifically: 

First, compared with prior HBM-PIM/GDDR-PIM architectures~\cite{he2025papi,heo2024neupims,park2024attacc,gu2025cent}, 
the DIMM-PIM architecture features \textit{multiple distributed, cooperating DRAM chips}, which aggravate two synchronization overheads:
(1) \textit{Data synchronization:} attention computation distributed across multiple DRAM chips necessitates data synchronization via cross-chip transfers, e.g., accumulating partial outputs across chips for ``softmax''.
(2) \textit{Layout synchronization:} the DIMM-based data storage layout may not match the layout required by PIM computation: 
for instance, a single element may be striped across several chips, while PIM computation requires each element resides in a single chip. 
Aligning the layout thus becomes a prerequisite for attention execution.
These overheads introduce substantial stalls in PIM execution, preventing DIMM-PIM from achieving its theoretical performance.

Second, the disaggregation of the inference process between the GPU and DIMM-PIM also introduces two types of synchronization overhead:
(1) \textit{Data synchronization:} PIM needs to fetch the Q,K,V tensors from the GPU and return the attention outputs to the GPU. 
These transfers via PCIe can stall PIM execution.
(2) \textit{Progress synchronization:} when parallel operations on the PIM and GPU have imbalanced completion times, the device that finishes earlier is forced to wait, creating idle ``bubbles''.
This issue is compounded by the fact that the execution latencies of parallelizable tasks can vary, making workload alignment more challenging.

To address these challenges, 
we propose \sys, the first AFD system that integrates a novel DIMM-PIM design as the accelerator, 
reducing synchronization overheads from the hardware internals to the overall system.
First, we design \sys-PIM (\textsection\ref{sec:design-dp}), a DIMM-PIM hardware for efficient attention computation.
It eliminates the overheads of data and layout synchronization with two key techniques respectively: \emph{bubble-free pipelining} and \emph{hybrid-grained re-layout}.
Specifically, the pipeline overlaps data transfer with concurrent bank PU execution to hide the transfer overheads, 
which further enables bubble-free with specific head mapping according to a quantitative analysis.
The hybrid-grained re-layout performs data layout transformation at element level (coarse-grained) and bit level (fine-grained) to address the layout mismatch with minimum latency.

Second, we design \sys-sys (\textsection\ref{sec:design-sys}), a DIMM-PIM integrated inference system that reduces the overheads of data and progress synchronization respectively with \emph{rankset-granular communication computation overlapping} and \emph{alignment-predicting scheduling}.
Rankset-granular communication computation overlapping exploits the finest granularity of independent communication and computation, i.e., \emph{the rankset}, which consists of one rank from each channel.
This design achieves parallel communication and computation across ranksets to support asynchronous data transfer, effectively hiding the communication overhead between the GPU and \sys-PIM with computation.
Alignment-predicting scheduling selects requests to form sub-batches with the ability of modeling the performance of operations on both devices, 
selecting requests accordingly to form sub-batches whose predicted latencies on the two devices are aligned.
This maximizes the parallel execution on the two devices with minimum idle bubbles.

Comprehensive evaluations on three LLM models and three real-world traces~\cite{open-r1,dophin,OpenThoughts-114k-math,bai2024longbench} show that \sys achieves up to 5.15$\times$ speedup over state-of-the-art HBM-PIM solutions~\cite{heo2024neupims,park2024attacc} with significantly improved batch sizes.

\begin{figure}[t]
   \setlength{\abovecaptionskip}{1pt}
   \setlength{\belowcaptionskip}{-10pt}
  \centering
   \begin{minipage}[t]{\linewidth}
    \centering
    \includegraphics[width=\textwidth]{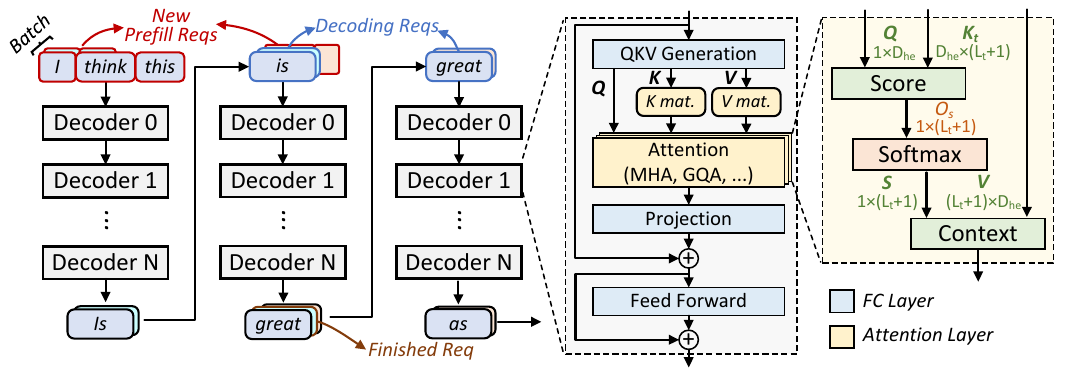}
    \footnotesize
    \end{minipage}
    \caption{\textbf{LLM inference process.} %
    }

 \label{fig:motiv-LLM-inference}
\end{figure}

\section{The Rise of AF-Disaggregated Systems}

\subsection{LLM Inference Process}
Fig.~\ref{fig:motiv-LLM-inference} illustrates the LLM inference process~\cite{anil2023palm2,devlin2019bert,grattafiori2024llama3,kalyan2023survey}, which includes two stages, i.e., prefilling and decoding. 
The LLM model constitutes a sequence of layers, 
each comprising two major kinds of operations: 
(1) attention operations~\cite{vaswani2017attention} and
(2) fully-connected (FC) operations, which encompass QKV Gen, projection, Feed-forward Network, etc. 
MHA and GQA are leading attention operators, with MHA computing each head independently and GQA grouping query heads for parallel processing.
The computations of attention consist of independent heads,
each following score ($Q \times K_t$), softmax, and context computations ($S \times V$).

\subsection{Attention-FC Disaggregated Inference System}
\label{sec:attn-fc-disaggregation}
Attention and FC operations typically have different characteristics. 
Specifically, benefiting from batching~\cite{amey2024sarathi,yu2022orca,kwon2023pagedattention}, FC operations are compute-intensive and demand compute resources for acceleration.
On the contrary, decoding attention is typically bandwidth-intensive~\cite{park2024attacc} and does not benefit from batching, since the KV cache is distinct and specific for each request. 
The latency of attention is primarily determined by memory bandwidth for loading the KV cache.
Considering the different characteristics of the two operations, 
some current works, i.e., Attention-FC disaggregated inference systems (``AFD systems'' in short), disaggregate the attention and FC operations on different hardware platforms.
Specifically, they batch FC operations on the compute-rich GPU/NPU, while offloading the KV cache and the decoding attention to a platform (called \textit{accelerators}) better suited for its memory-intensive nature.

\begin{figure}[t]
  \setlength{\abovecaptionskip}{1pt}
  \setlength{\belowcaptionskip}{-10pt}
 \centering
  \begin{minipage}[t]{\linewidth}
   \centering
   \includegraphics[width=\textwidth]{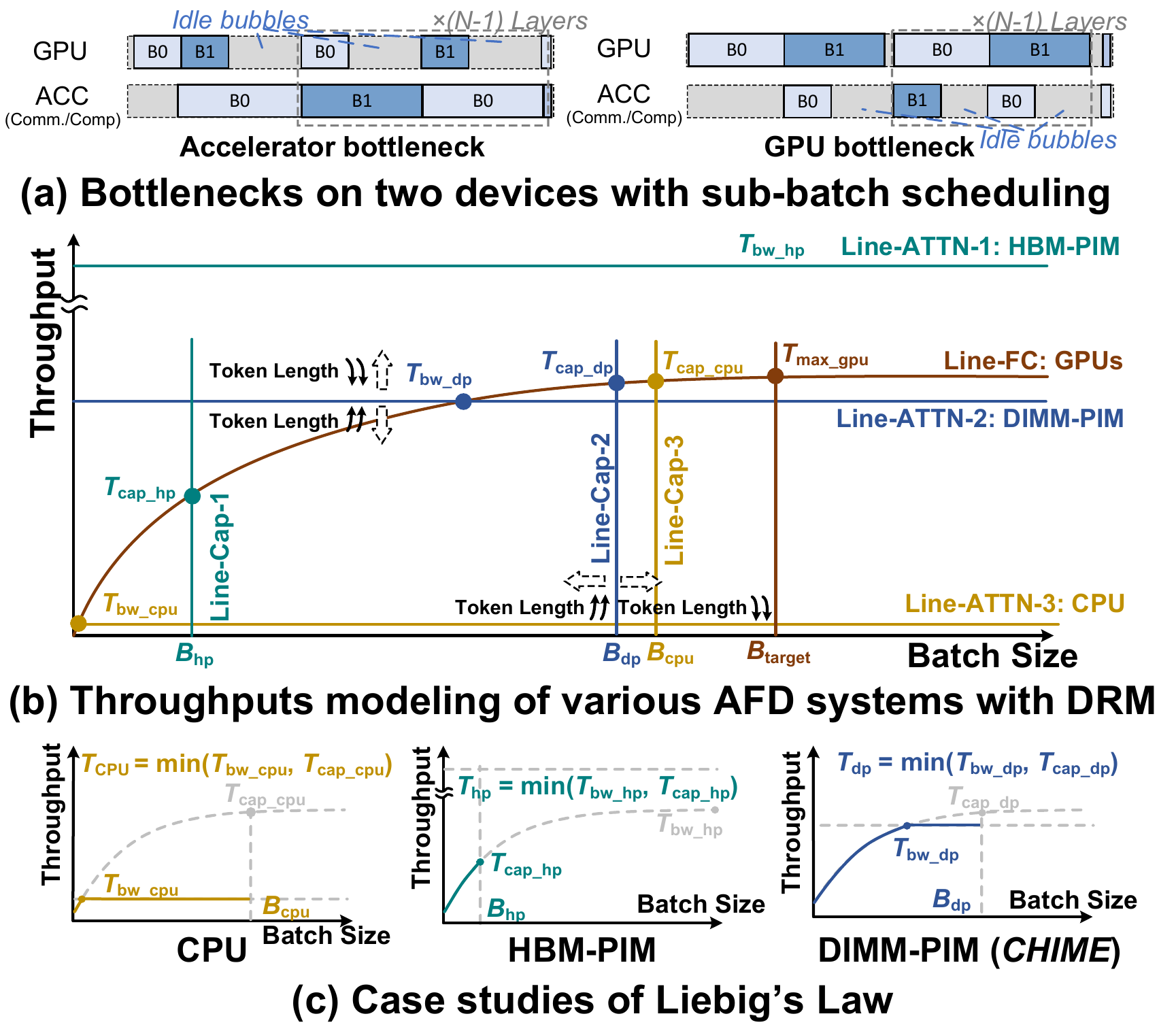}
   \footnotesize
   \end{minipage}
   \caption{\textbf{\sys's Disaggregated Roofline Model and case studies.} 
        \textit{``DIMM-PIM'' denotes \sys's proposal, leveraging DIMM-based host memory with bank-level PUs as the accelerator. ``Throughput'' denotes tokens per second.}
   }
 \label{fig:intro-motiv}
\end{figure}

To improve the overall utilization, 
current AFD systems typically apply \emph{sub-batch scheduling}~\cite{heo2024neupims,jiang2024neo,zhu2025megascaleinfer}, 
making two (or even more) interleaved sub-batches to run in parallel across the two devices, as shown in Fig.~\ref{fig:intro-motiv}-a.
During inference,
the execution of an individual batch cannot be parallelized on two devices, e.g., GPU has to wait for the completion of attention before starting FC.
Sub-batch scheduling enables attention and FC of different batches to run concurrently on the two devices, which could improve the throughput and resource utilization of AFD systems.

\subsection{Processing-In-Memory (PIM)}
PIM is a promising solution to offer high aggregated bandwidth by integrating processing units (PUs) in memory devices, such as HBM~\cite{lee2021hbmpim,he2020newton,kim2021aquabolt,kim2021aquabolthcs,lee2025paise}, 
GDDR~\cite{lee2022gddr6,kwon2022system,kim2024skhynix,kwon2023memorycentric,lee2024cost}, 
and DIMM~\cite{lee2022axdimm,devaux2019true,ke2020recnmp,ke2021near}.
For example, with two DIMMs (e.g., two ranks of 16 banks per DIMM) in one channel, 
placing PUs near ranks (\textbf{rank PU}) can achieve about 4$\times$ the bandwidth of host CPU, while integrating PUs near banks (\textbf{bank PU}) can improve bandwidth by more than 30$\times$. 
In this case, some state-of-the-art AFD systems are built upon PIM devices, such as  HBM-PIM~\cite{heo2024neupims,park2024attacc,he2025papi} or GDDR-PIM~\cite{gu2025cent}, delivering substantial performance gains over GPUs.

\section{Characterizing AFD Systems with DRM}
\label{s:demys}
In this section, we first introduce \textbf{Disaggregated Roofline Model (DRM)}, a unique analytical model specifically designed for AFD systems, characterizing the performance across diverse scenarios.
Based on the model, we analyze the memory bottlenecks for current AFD systems whose attention is offloaded to various memory-enhanced devices, 
including HBM-PIM, which boosts bandwidth, and CPUs (with host memory), which expand capacity.
According to the analysis, we conclude how the ``Liebig's Law'' manifests for AFD systems, 
which informs the design of accelerators with effective memory enhancement to achieve high throughput.

\subsection{DRM Modeling and Bottleneck Analysis}
\label{s:motiv-analytic-model}
Characterizing the throughput of AFD systems requires comprehensively considering two factors: 
the throughput on the GPU side for executing FC operations, 
and the throughput on the accelerator side for executing attention operations.
We first identify that when the throughputs on the two devices are not equal, the throughput of the inference system is only affected by the device that becomes the bottleneck.
For example in the right of Fig.~\ref{fig:intro-motiv}-a, the bottleneck is the GPU, and the accelerator suffers idle bubbles consequently.
In this case, increasing the throughput on the accelerator side will only result in more idle bubbles, having no effect on improving the overall throughput.
The left of Fig.~\ref{fig:intro-motiv}-a shows an opposite example.
In conclusion:

\begin{mdframed}[backgroundcolor=gray!10,linecolor=black,linewidth=1pt]
\textbf{Implication~\uppercase\expandafter{\romannumeral1}} 
\textit{The overall inference throughput of AFD systems is determined by the lower throughputs of the two devices that separately execute the FC and attention.
}
\end{mdframed}

\myparagraph{Construction of DRM.}
\REVISE{
Fig.~\ref{fig:intro-motiv}-b presents a conceptual illustration of the DRM,
which uniformly characterizes the performance of two disaggregated devices by modeling the relationship between batch size and token throughput (tokens per second).
Token throughput is derived from two key metrics:
(1) floating point operations per second, which is dictated by the batch-size-dependent arithmetic intensity based on traditional roofline model.
(2) floating point operations, which scales with the batch size.
Line-FC shows the GPU token throughput for executing the FC, 
which increases with the growth of batch size until the GPU is fully utilized, i.e., when the batch size reaches $B_{\rm target}$, the GPU achieves peak throughput $T_{\rm max\_gpu}$.
Line-ATTN-1, 2, 3 show the token throughputs of different accelerators for executing the attention respectively.
Since the arithmetic intensity of attention remains constant and low, 
these throughputs are typically positively correlated with accelerator memory bandwidth, regardless of batch size.
Moreover, The performance curves in Fig.~\ref{fig:intro-motiv}-b can be not only theoretical, but also profiled.
}

\myparagraph{Bandwidth constraints.}
According to Implication~\uppercase\expandafter{\romannumeral1}, the throughput of an AFD system is determined by the lower value between the throughput curves of the GPU and the accelerator.
There can be two types of situations:
First, the throughput curves of the GPU and the accelerator intersect, e.g., when the accelerator is the CPU. %
The GPU throughput may initially be lower than that of the CPU when the batch size is very low, and the throughput of the AFD system is determined by Line-FC.
As the batch size increases, the GPU throughput quickly rises to $T_{\rm bw\_cpu}$, after which the throughput of the AFD system is determined by Line-ATTN-3, which is constrained by the CPU memory bandwidth.
Second, if the throughput curves of the GPU and the accelerator do not intersect, the throughput of the AFD system is always determined by the lower curve. 
E.g., if the accelerator is HBM-PIM in Fig.~\ref{fig:intro-motiv}-b, the throughput is always limited by the GPU side (Line-FC).

\begin{mdframed}[backgroundcolor=gray!10,linecolor=black,linewidth=1pt]
\textbf{Implication~\uppercase\expandafter{\romannumeral2}} 
\textit{The bandwidth of the accelerator limits the throughput of AFD systems when the attention becomes the bottleneck.
}
\end{mdframed}

\myparagraph{Capacity constraint.}
Another factor is the memory capacity on the accelerator side to store the KV cache. 
Insufficient memory capacity restricts the maximum number of requests that can be accommodated, preventing the GPU side from fully utilizing GPU resources. 
In Fig.~\ref{fig:intro-motiv}-b, Line-Cap-1, 2, 3 show the maximum batch sizes that can be accommodated by various accelerators, i.e., $B_{\rm hp}$, $B_{\rm dp}$, $B_{\rm cpu}$ for HBM-PIM, DIMM-PIM and CPU respectively.
The GPU throughput for executing the FC is thus limited by these maximum batch sizes.
For example, when the accelerator is the HBM-PIM, the maximum GPU throughput on Line-FC is limited below $T_{\rm cap\_hp}$, since the batch size of this AFD system is at most $B_{\rm hp}$.
When the bottleneck is on the GPU side, e.g., HBM-PIM as the accelerator, the AFD system's throughput is constrained by the limited batch size.

\begin{mdframed}[backgroundcolor=gray!10,linecolor=black,linewidth=1pt]
\textbf{Implication~\uppercase\expandafter{\romannumeral3}} 
\textit{The capacity that stores the KV cache limits the overall throughput of AFD systems when the FC becomes the bottleneck.
}
\end{mdframed}

\myparagraph{Dynamic workloads and shifting rooflines.}
As DRM is an analytical model for design-time architectural analysis, 
its purpose is not to predict the precise latency of a given request~\cite{stojkovic2025dynamoLLM,kakolyris2025throttllem}. 
Nevertheless, DRM could also illustrate how the performance rooflines shift in response to varying runtime workload characteristics.
For example, with longer context, the maximum batch size that can be accommodated within the same capacity and the throughput of processing a larger KV cache with the same bandwidth decreases, 
so it causes the capacity lines (Line-Cap-1, 2, 3 in Fig.~\ref{fig:intro-motiv}-b) to shift leftward and the Line-ATTN-1, 2, 3 to shift downward, 
while for the GPU, Line-FC remains almost unchanged. 
Consequently, it illustrates how longer contexts exacerbate memory bottlenecks along both the bandwidth and capacity.

\myparagraph{\REVISE{Characterizing communication overhead.}}
\REVISE{
The methodology of DRM is also capable of characterizing the communication overhead between GPUs and attention accelerators,
considering it as a part of attention overhead.
Within the DRM, the decrease of communication bandwidth (e.g., utilizing PCIe 3.0 instead of PCIe 4.0)
results in a downward shift of the attention performance curve.
According to our detailed analysis in \textsection\ref{sec:design-sys-comm}, the communication overhead in AFD systems is constant, and is optimized by our design.
}

\myparagraph{The ``Liebig's Law'' for AFD systems.}
Considering Implication~\uppercase\expandafter{\romannumeral2} and~\uppercase\expandafter{\romannumeral3}, 
for a given GPU configuration that executes FC, 
the accelerator's memory bandwidth and capacity respectively limit the throughput when the attention or FC becomes the bottleneck. 
Further, according to Implication~\uppercase\expandafter{\romannumeral1}, this indicates that the overall throughput is determined by the memory capacity or bandwidth whose throughput limit is lower,
which we refer to as the weaker point of memory.
We conclude that the accelerator's memory impact on overall throughput follows Liebig's Law, that is:

\begin{mdframed}[backgroundcolor=gray!10,linecolor=black,linewidth=1pt]
\textbf{Liebig's Law:}
\textit{The throughput of an AFD system is constrained by the weaker point in terms of either memory capacity or bandwidth.
}
\end{mdframed}
Although Liebig's Law is a general principle applicable to any system, the analysis methodology with DRM first reveals how the Law manifests for AFD systems.

\myparagraph{Case study.} 
Fig.~\ref{fig:intro-motiv}-c extracts concrete examples from Fig.~\ref{fig:intro-motiv}-b as case studies. 
The weaker point of the ``CPU'' is the limited bandwidth of the host memory, 
so the throughput of the CPU AFD system is constrained by $T_{\rm \text{bw\_cpu}}$ even if it has sufficient memory capacity to accommodate larger batch sizes. 
The weaker point of the ``HBM-PIM'' is the limited HBM capacity, so the throughput is constrained by $T_{\rm \text{cap\_hp}}$ even if it can execute attention extremely quickly.

\begin{figure}[t]
  \setlength{\belowcaptionskip}{-10pt}
 \centering
 \begin{minipage}[t]{0.96\linewidth}
  \centering
  \includegraphics[width=\textwidth]{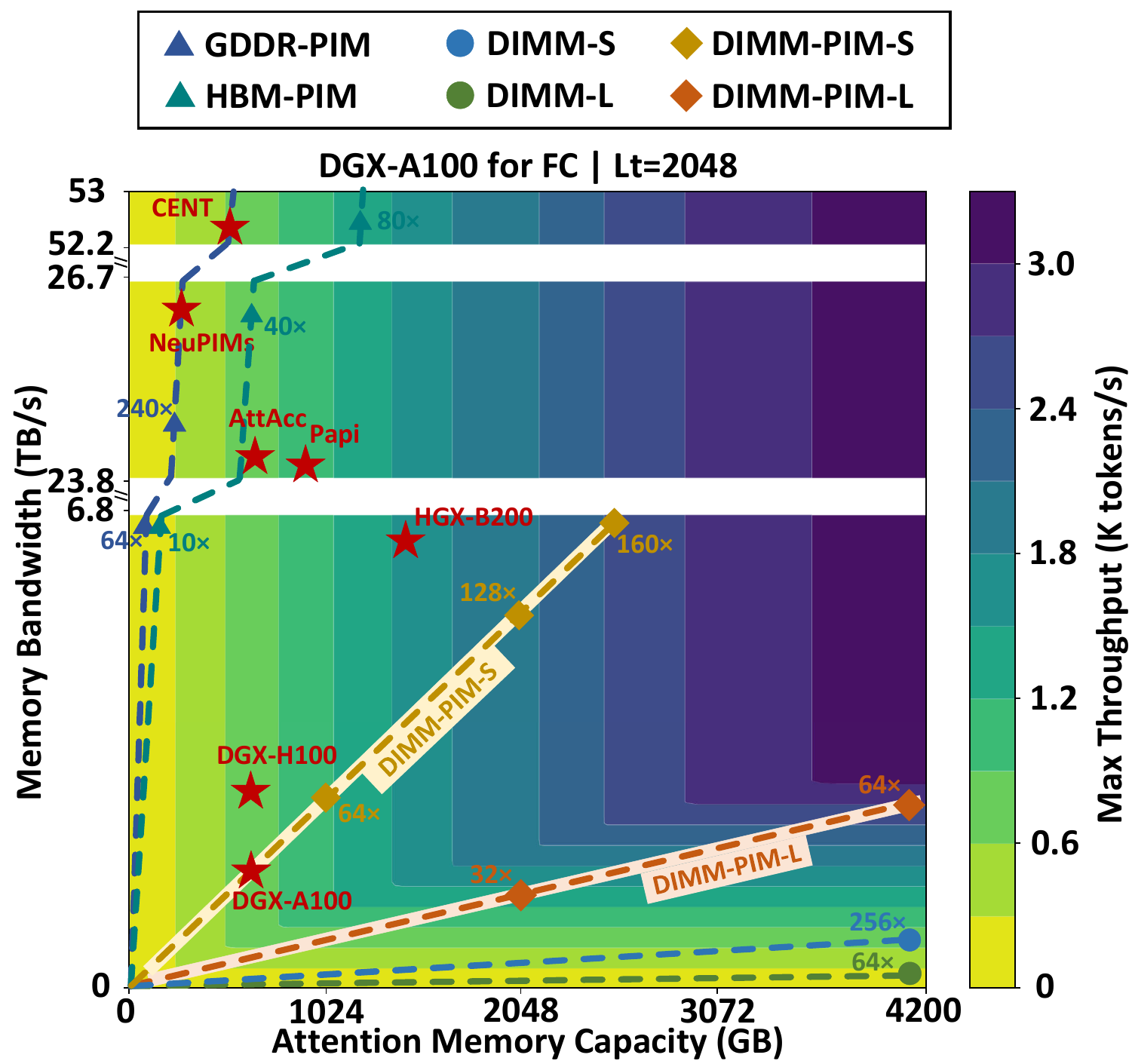}
  \footnotesize
  \end{minipage}
  
  \caption{\REVISE{\textbf{Contour plot of the throughput with memory bottlenecks.}} 
\textit{
\REVISE{
``$L_{t}$=2048'' above denotes that each request has a context length of 2048 tokens.
Memory configurations of ``PAPI'', ``NeuPIMs'', ``AttAcc'' and ``CENT'' are derived from the papers~\cite{he2025papi,heo2024neupims,park2024attacc,gu2025cent}.
``DGX-A100'', ``DGX-H100'' and ``HGX-B200''~\cite{nv2023dgxa100,hgxb200,h20} represents using GPUs for attention.
DIMM-S, DIMM-L denote two different types of DDR4 memory.
Every point represents the total memory capacity and bandwidth under the corresponding number of memory chips, 
for example, 
the point ``64$\times$'' on the ``DIMM-L'' line (green line) represents the memory capacity and bandwidth corresponding to 64$\times$ DIMM-L memory chips.
We assume that the PIM performance of each memory device can reach the ideal case, i.e., the equivalent memory bandwidth is scaled by 16$\times$.
The performance is simulated with AttAcc~\cite{park2024attacc} simulator with GPT-175B model.
}}
   }
\label{fig:motiv-memory-bottleneck}
\end{figure}

\myparagraph{\REVISE{Quantitative analysis on Liebig's Law.}}
\REVISE{
To illustrate Liebig's Law quantitatively, we extend the illustrative figure of Fig.~\ref{fig:intro-motiv} to Fig.~\ref{fig:motiv-memory-bottleneck}.
Fig.~\ref{fig:motiv-memory-bottleneck} quantifies the throughput limitation assuming 8$\times$A100 of DGX-A100 are used for computing FC, 
while using accelerators for attention with various memory configurations.
It clearly demonstrates Liebig's Law, i.e., better throughput can only be achieved when both capacity and bandwidth are scaled simultaneously. 
If only one of them is scaled, for example, fixing the capacity and scaling the bandwidth, 
the throughput will only improve with increasing bandwidth until the bandwidth reaches a certain point. 
Beyond that point, higher bandwidth yields no further improvement, as capacity becomes the bottleneck.
}

\subsection{Limitations on Prior Works}
Existing works have made considerable efforts to enhance memory for improving inference throughput.
However, they mainly focus on addressing one dimension of the memory bottleneck, either capacity or bandwidth.
Their neglect of Liebig's Law renders the resulting performance gains potentially inefficient or insufficient. 
\REVISE{
To illustrate, we plot the hardware configurations of various hardware types under scaling scenarios in Fig.~\ref{fig:motiv-memory-bottleneck}:
including HBM2e, GDDR, two types of DDR4 memory and their PIM variants.
Some listed systems~\cite{gu2025cent} in Fig.~\ref{fig:motiv-memory-bottleneck} are not specifically designed for AFD, and we mainly consider their hardware for executing attention.
We can conclude that prior works suffer from the ``Liebig's Law'' in the following aspects:
}

\myparagraph{Insufficient memory enhancement.}
\REVISE{
First, without adequate scaling, 
these memory enhancement techniques could be insufficient to prevent memory from becoming a throughput bottleneck. 
This can be illustrated by examining the hardware parameters claimed in their respective papers in conjunction with Fig.~\ref{fig:motiv-memory-bottleneck}. 
Specifically, works based on HBM-PIM could suffer from insufficient memory capacity~\cite{he2025papi,heo2024neupims,park2024attacc}, 
while others suffer from insufficient memory bandwidth of DIMMs~\cite{liu2025hermes,jiang2024neo}.
}

\myparagraph{Inefficient memory enhancement.}
\REVISE{
Further, current works suffer from excessive memory resources. 
For example, although current GDDR/HBM-PIM based methods~\cite{he2025papi,heo2024neupims,park2024attacc,gu2025cent} are limited by capacity, 
their bandwidth could be orders of magnitude larger than the requirement, which provides almost no benefit to throughput. 
Moreover, even if scaling them could potentially achieve high throughput, 
such scaling could simultaneously lead to substantial bandwidth waste, 
as illustrated in Fig.~\ref{fig:motiv-memory-bottleneck}.
}

\subsection{Proposal of Choosing DIMM for PIM Intergration}
\label{sec:motiv-dimm-pim-proposal}
Based on our analysis, we find that integrating PIM with \textit{DIMM}s (i.e., \textit{DIMM-PIM}) is a promising accelerator for attention. 
Specifically, \textit{DIMM-PIM} augments DIMM-based host memory with bank-level PIM units, combining the strengths of PIM and \textit{DIMM}s. 
Therefore, we propose AFD systems that integrate \textit{DIMM-PIM}, which offer the following benefits:

\myparagraph{Better performance with more balanced memory configuration.}
\REVISE{
Typically, DIMM-PIM can alleviate the capacity bottleneck of HBM-PIM and the bandwidth bottleneck of CPU offloading.
As illustrated in Fig.~\ref{fig:intro-motiv}-b, c, this means that DIMM-PIM potentially improves the overall throughput with a more balanced memory configuration.
As shown in Fig.~\ref{fig:motiv-memory-bottleneck}, since the DGX-A100 is equipped with 2TB of DIMM memory (DDR4) and 640GB of HBM memory, 
equipping the DIMMs with PIMs already significantly outperforms equipping the HBMs with PIMs.
}

\myparagraph{Scalable and configurable.}
DIMM-PIM offers scalability via the standard DIMM form factor, 
enabling plug-and-play capacity/bandwidth scaling on compatible platforms, 
and can be further scaled using interconnects such as CXL.
This allows DIMM-PIM to not only achieve better performance but also potentially adapt to a broader range of scenarios.

\myparagraph{Economic efficiency.}
\REVISE{
Beyond performance considerations, DIMM-PIM also offers a significant cost advantage 
over GDDR- and HBM-based alternatives.
First, the per-GB cost of HBM2e can be over $6\times$ higher than DDR4~\cite{cost-ddr4,cost-ddr5,cost-ddr5-hbm}, which represents a substantial difference, regardless of whether PIM is integrated. 
Second, Fig.~\ref{fig:motiv-memory-bottleneck} shows that, 
the more balanced performance of DIMM-PIM avoids the significant resource over-provisioning inherent in HBM/GDDR-PIM solutions,
enabling more cost-effective performance scaling.
}

\begin{figure}[t]
  \setlength{\belowcaptionskip}{-10pt}
 \centering
 \begin{minipage}[t]{0.48\linewidth}
  \centering
  \includegraphics[width=\textwidth]{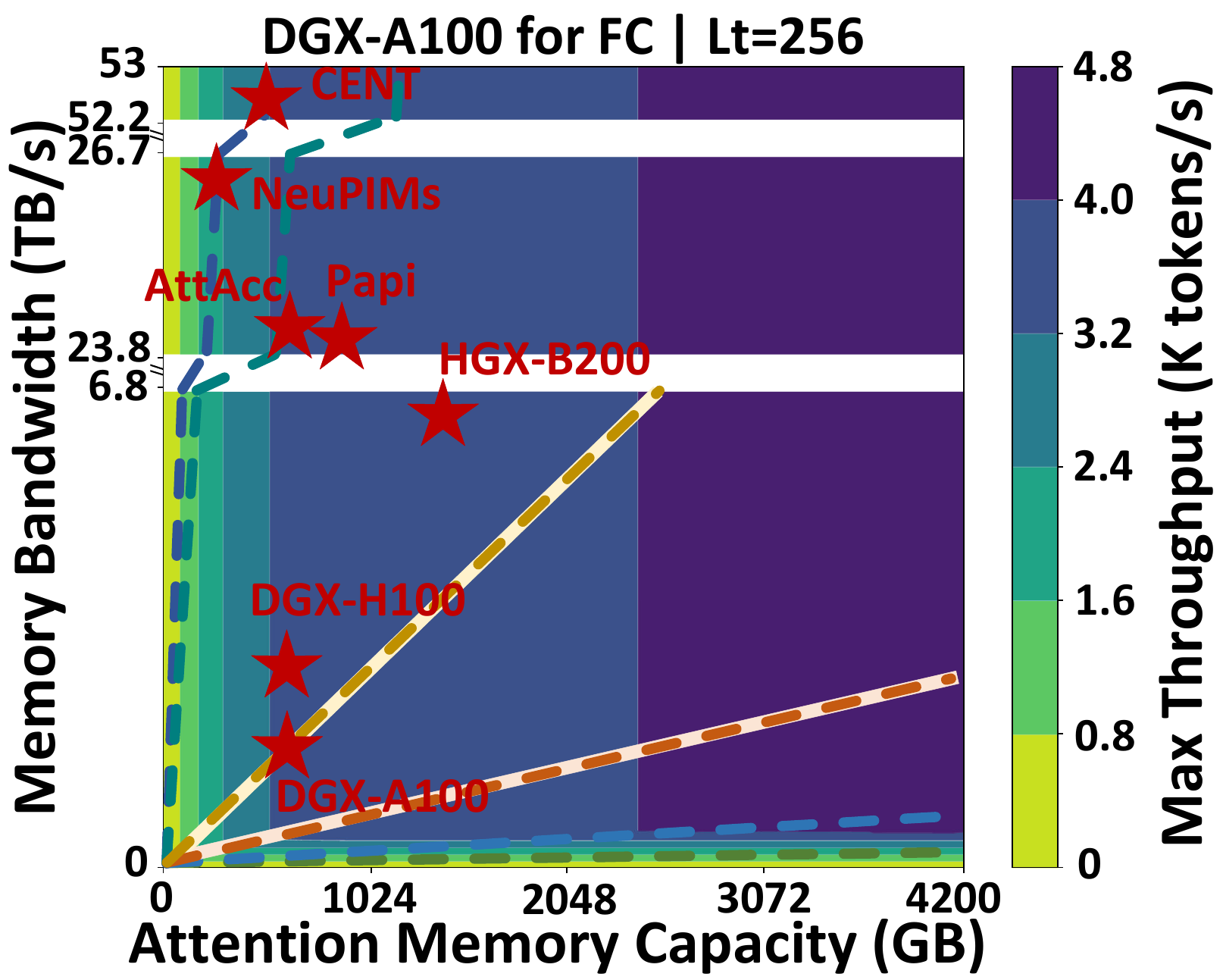}
  \footnotesize
   \textbf{\REVISE{(a) $L_{t}$=256.}}
  \end{minipage}
  \begin{minipage}[t]{0.48\linewidth}
    \centering
    \includegraphics[width=\textwidth]{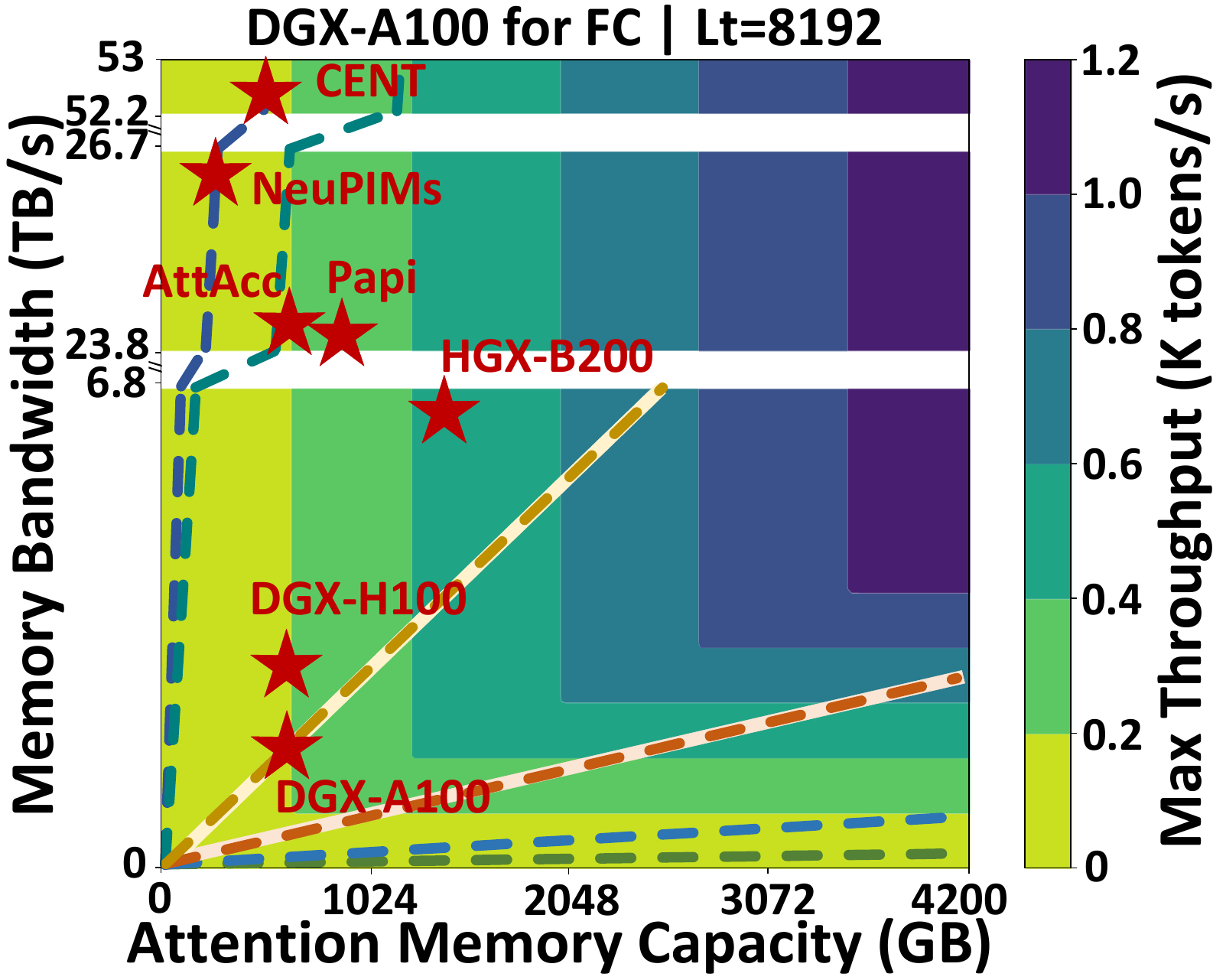}
    \footnotesize
     \textbf{\REVISE{(b) $L_{t}$=8192.}}
  \end{minipage}
  \begin{minipage}[t]{0.48\linewidth}
    \centering
    \includegraphics[width=\textwidth]{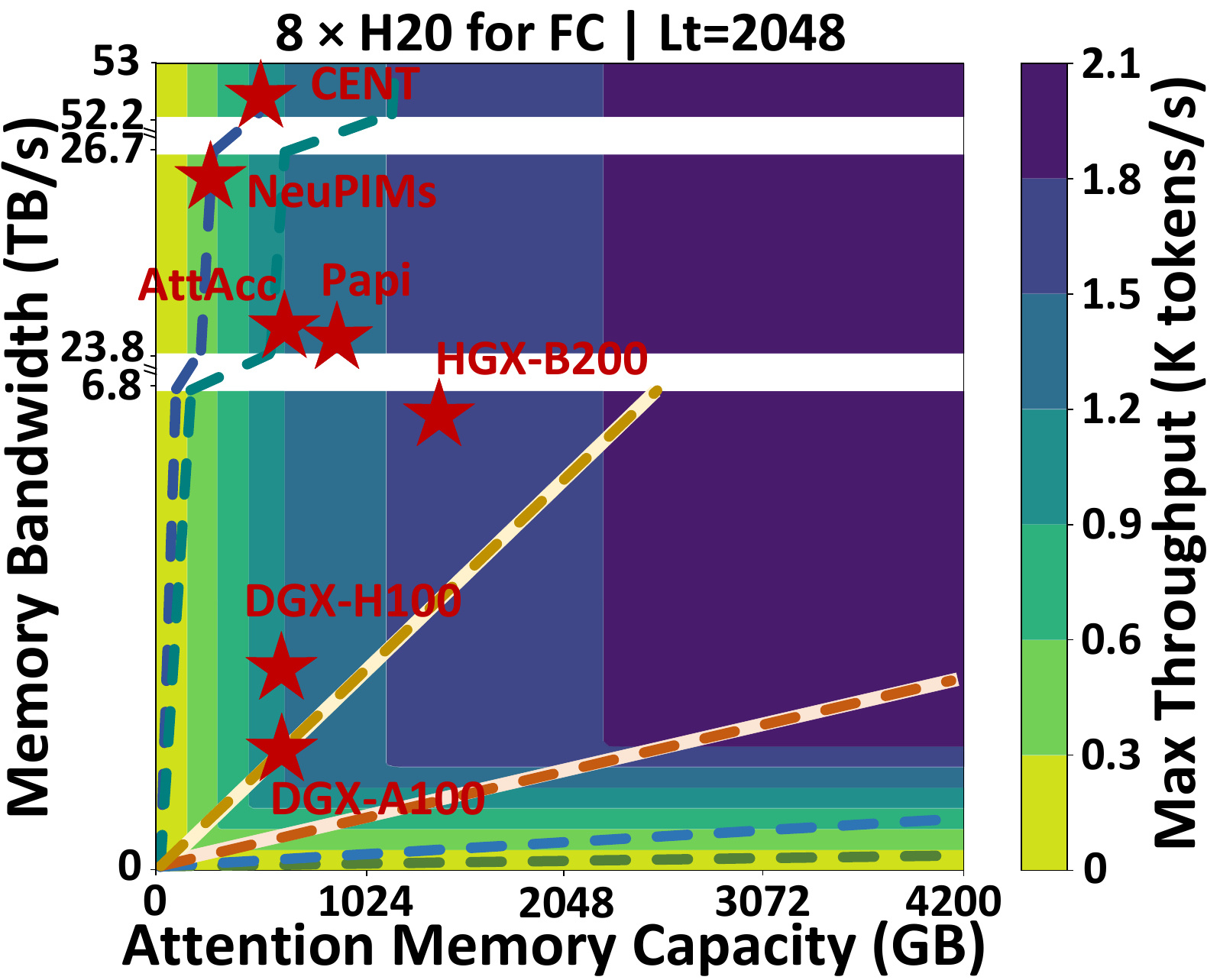}
    \footnotesize
      \textbf{\REVISE{(c) 8$\times$H20 for FC.}}
  \end{minipage}
  \begin{minipage}[t]{0.48\linewidth}
    \centering
    \includegraphics[width=\textwidth]{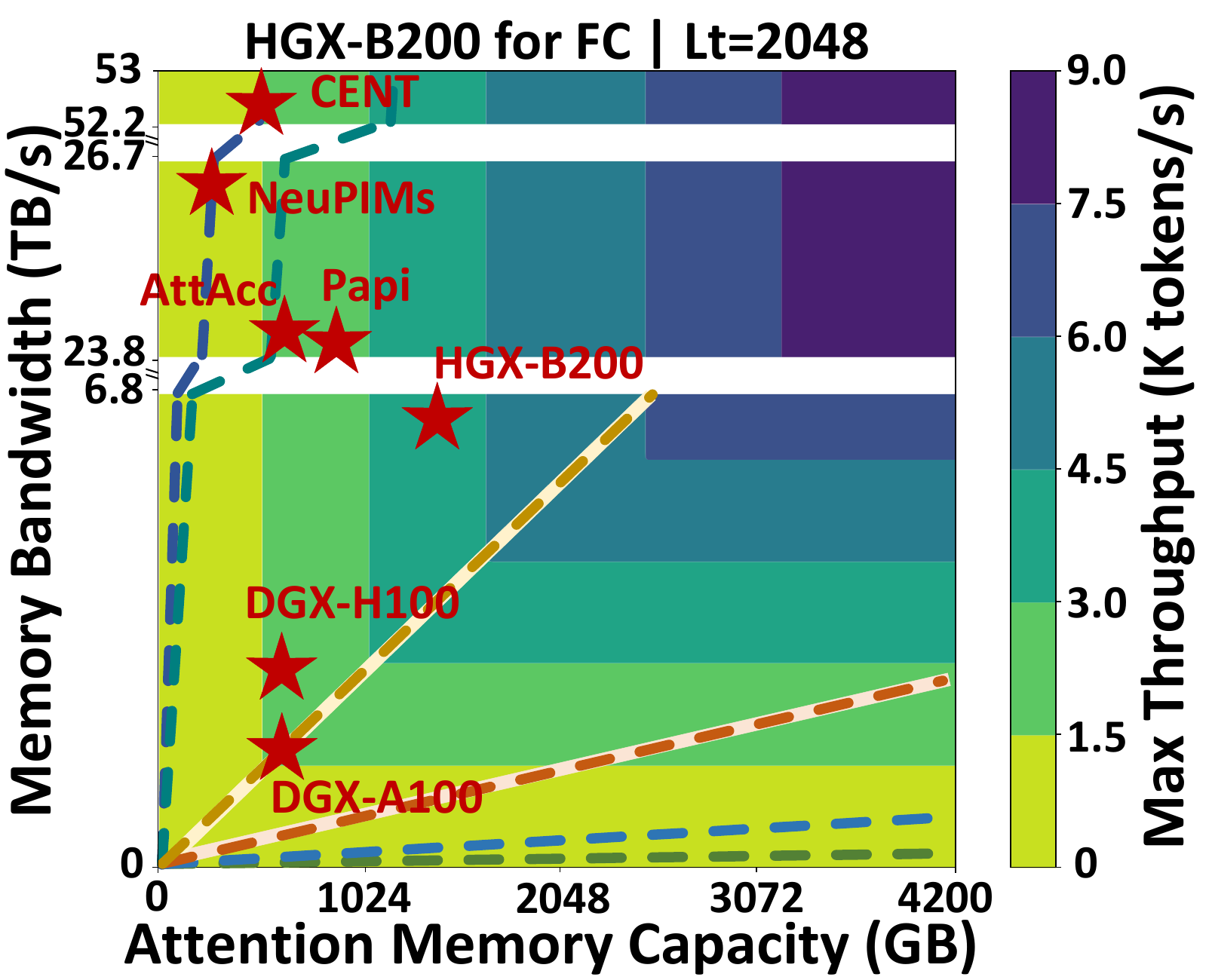}
    \footnotesize
      \textbf{\REVISE{(d) HGX-B200 for FC.}}
  \end{minipage}
  
  \caption{\textbf{\REVISE{Contour plot of the throughput for four scenarios.}}
        \textit{\REVISE{This figure shares the legend with Fig.~\ref{fig:motiv-memory-bottleneck}. 
        Compared to the scenario in Fig.~\ref{fig:motiv-memory-bottleneck}, 
        (a) and (b) vary the context lengths, 
        while (c) and (d) adjust the number of GPUs used for FC computation.}  }
   }
\label{fig:motiv-scenario-shifting}
\end{figure}

\begin{figure}[t]
  \setlength{\belowcaptionskip}{-10pt}
 \centering
 \begin{minipage}[t]{0.96\linewidth}
  \centering
  \includegraphics[width=\textwidth]{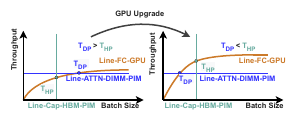}
  \footnotesize
  \end{minipage}
  \caption{\REVISE{\textbf{DRM with GPU upgrade.} 
  \textit{$T_{DP}$ and $T_{HP}$ denote the throughput of DIMM-PIM under its memory bandwidth constraint, and the throughput of HBM-PIM under its memory capacity constraint, respectively.}}}
\label{fig:motiv-gpu-upgrade}
\end{figure}

\subsection{Discussion: Bottleneck Shifting across Various Scenarios.}
\label{sec:motiv-discussion-scenarios}
\REVISE{
In practice, AFD systems may adopt various deployment configurations, which could shift the memory bottleneck.
The deployment configurations include GPU model and quantity, context length, LLM parameter size and architecture, etc. 
An important contribution of our DRM-based methodology is its ability to analyze, for a given deployment configuration, how the memory configuration of each accelerator affects overall throughput.
In this section, we further discuss how different deployment configurations could affect the memory bottleneck.
}

\myparagraph{Different context lengths.}
\REVISE{
Fig.~\ref{fig:motiv-memory-bottleneck} and Fig.~\ref{fig:motiv-scenario-shifting}-a, b show how different context lengths affect the performance.
First, we observe that across various context lengths, 
the more balanced configuration of DIMM-PIM consistently holds a performance and efficiency advantage, 
effectively improving throughput while reducing resource over-provisioning. 
It is because longer contexts exacerbate memory capacity and bandwidth bottlenecks simultaneously. 
}

\REVISE{
Second, we observe that with shorter contexts, 
the performances of various baselines become more comparable, 
because the marginal benefit of adding resources diminishes when the memory bottleneck is alleviated. 
We validate this analysis in \textsection\ref{sec:eval-e2e-throughputs}. 
The impact of context length can also be analyzed in conjunction with Fig.~\ref{fig:intro-motiv}-b.
}

\myparagraph{Influence of other configurations.}
\REVISE{
Our DRM method can also analyze the impact of other configuration changes on accelerator selection. 
For example, with more advanced GPUs for executing FC, as shown in Fig.~\ref{fig:motiv-gpu-upgrade}, 
the DRM reveals that this shifts the curve of FC throughput upward, 
enhancing HBM-PIM throughput while leaving DIMM-PIM throughput unchanged, 
thereby closing the performance gap of HBM-PIM relative to DIMM-PIM and even turning it into an advantage.
Fig.~\ref{fig:motiv-memory-bottleneck} and Fig.~\ref{fig:motiv-scenario-shifting}-c, d show examples of this: 
32$\times$ DIMM-PIM-L outperforms AttAcc~\cite{park2024attacc} with A100 (Fig.~\ref{fig:motiv-memory-bottleneck}), but underperforms AttAcc when paired with B200 (Fig.~\ref{fig:motiv-scenario-shifting}-d).
Other scenarios, such as varying the number of GPUs used for FC, can also be analyzed following a similar methodology.
}

\section{Overview of \sys}

\subsection{Design Overview}
\begin{figure}[t]
  \setlength{\abovecaptionskip}{1.2pt}
  \setlength{\belowcaptionskip}{-15pt}
  \centering
    \begin{minipage}[t]{0.92\linewidth}
    \centering
    \includegraphics[width=\textwidth]{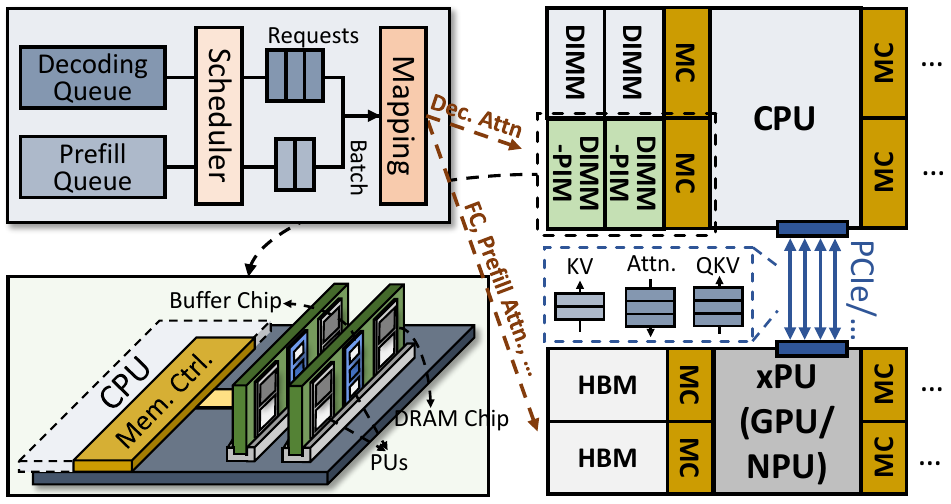}
    \footnotesize
    \end{minipage}
    \caption{\textbf{\sys overview.}}
  \label{fig:design-overview}
\end{figure}

To support the proposal of integrating DIMM-PIM, as shown in Fig.~\ref{fig:design-overview}, we propose \sys that consists of two parts: 
(1) \sys-PIM, \REVISE{our specifically designed DIMM-PIM hardware for efficient attention execution}; and
(2) \sys-sys, a scalable AFD system which coordinates the computation on GPU and \sys-PIM and aims at maximizing the inference throughput.

\myparagraph{Inference workflow of \sys.}
\sys-sys inherits the sub-batch scheduling technique~\cite{heo2024neupims,jiang2024neo}, selecting requests to form sub-batches for each iteration.
Initially, for each sub-batch, QKV Generation of all requests is batched and executed on the GPU. 
Then, prefilling and decoding attention are processed concurrently on the GPU and \sys-PIM respectively. 
\sys aggregates attention outputs from all requests in the batch, and then performs batched execution of subsequent operations (projection, Feed-forward, etc). 
PCIe interconnects the GPU and \sys-PIM.

\myparagraph{Addressing the challenges of integrating DIMM-PIM.}
As outlined in \textsection\ref{sec:intro}, integrating DIMM-PIM introduces architectural disaggregation and, 
consequently, synchronization overheads: 
within a DIMM, distributed chips incur both data-synchronization and layout-synchronization costs, while the disaggregation between DIMM-PIM and the GPU adds the overhead of cross-device data synchronization and progress synchronization. 
To address the former, \sys-PIM employs bubble-free pipelining (\textsection\ref{sec:design-dp-pipeline}) and hybrid-grained re-layout (\textsection\ref{sec:design-dp-relayout}); 
to address the latter, \sys-sys employs rankset-granular communication-computation overlapping (\textsection\ref{sec:design-sys-comm}) and alignment-predicting scheduling (\textsection\ref{sec:sys-scheduler}).

\section{Attention Acceleration on \sys-PIM} 
\label{sec:design-dp}

\begin{figure*}[t]
  \setlength{\abovecaptionskip}{-5pt}
  \setlength{\belowcaptionskip}{-10pt}
  \centering
    \begin{minipage}[t]{0.95\linewidth}
    \centering
    \includegraphics[width=\textwidth]{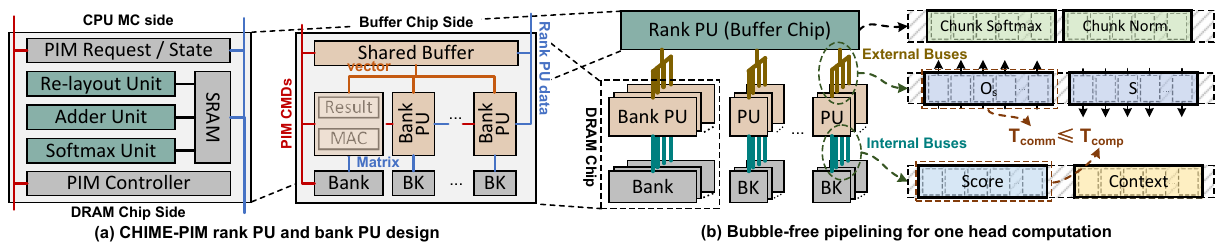}
    \footnotesize
    \end{minipage}
    \caption{\textbf{\sys-PIM design with bubble-free pipelining.}}
  \label{fig:design_arch_pipe}
\end{figure*}

This section describes \sys-PIM, a novel DIMM-PIM hardware design for efficient attention computation.

\myparagraph{Hardware components.}
As shown in Fig.~\ref{fig:design_arch_pipe}-a, \sys-PIM integrates co-operated processing units (PUs) across both bank and rank levels.
The bank PUs are integrated near DRAM banks with DRAM process, fetching KV cache from banks and performing score and context computation, 
while a shared buffer is leveraged to broadcast input vectors to all bank PUs. 
The bank PUs in all DRAM chips perform multiplication-and-accumulation (MAC) concurrently and generate outputs to the result buffer.
Softmax unit, adder unit, and re-layout unit are integrated on the buffer chip with logic process as a part of the rank PU, which can execute in an asynchronous manner with bank PUs.

\REVISE{The \sys-PIM workflow avoids modifications to host CPU memory controllers, which works as follows:
first, CPU offloads PIM requests to the rank PU via normal writes, which are further decoded to PIM commands.
Then, a dedicated PIM controller issues PIM commands on standard DDR interface to DRAM chips to trigger corresponding operations.
PIM commands include the ones for metadata setup and data movement, and some key commands are listed as follows. 
\texttt{PIM\_WR\_R} writes necessary information to registers, 
such as configurations of computing paradigm (adder tree or accumulator), self-incrementing index mechanisms of buffers, and so on.
\texttt{PIM\_LD\_SB} loads data from a certain bank to the shared buffer, while \texttt{PIM\_WR\_SB} writes data from rank PU to the shared buffer.
\texttt{PIM\_MAC} executes MAC operations in all banks while reading necessary data from DRAM cells and shared buffer.
\texttt{PIM\_RD\_RB} loads data from result buffers of all banks to rank PU. 
}

\subsection{Bubble-free pipelining}
To eliminate data-synchronization overhead in attention computation, 
the key is to leverage DIMM-PIM's unique architecture to design a pipelined attention execution that hides communication behind computation. 
Furthermore, through quantitative analysis and a tailored head-mapping method, we ensure that the communication cost is completely hidden.

\myparagraph{Orchestrating the pipeline with decoupled memory buses and kernel fusion.}
We orchestrate a pipelined execution for attention kernels in PIM that aggressively overlaps computation and communication, grounded in two architectural observations. 
First, the internal memory buses (servicing bank PUs accessing banks) and external memory buses (handling rank PUs accessing bank PUs or shared buffer) can be decoupled~\cite{chen2025asyncdimm}, 
enabling simultaneous execution of compute kernels on bank PUs and data transfer from/to rank PU, as shown in Fig.~\ref{fig:design_arch_pipe}-b.
Second, inspired by FlashAttention's kernel fusion technique~\cite{dao2022flashatten} (where attention is computed in chunked tiles),
we orchestrate fine-grained kernel pipelining to maximize concurrency while constraining the intermediate head footprint on rank PUs.

Concretely, during the scoring phase, 
when each bank PU computes a token-specific output $O_s$ and stages it in its local result buffer,
the rank PU immediately fetches it via otherwise-idle external buses.
Our attention kernel operates at the granularity of chunked tiles, where each chunk consists of data generated in parallel by a single head across (possibly) multiple bank PUs.
After accumulation in adder unit,
the softmax unit applies per-chunk softmax operations.
In this way, score and softmax computations are pipelined between bank PUs and rank PU in a chunk-based manner.
Following the processing of all tokens, a final cross-chunk normalization pass produces the globally-correct softmax output $S$, 
which is also performed in a streaming, chunk-wise manner.
It allows the computation of finalized $S$ elements, the write-back communication of $S$ to DRAM chips, and subsequent context computations to be pipelined.

\myparagraph{Enabling bubble-free with quantitative analysis and specific head mapping.}
\label{sec:design-dp-pipeline}
Due to the nature of DIMM that multiple co-operated DRAM chips form a rank,
KV matrix (cache) and related computation would be distributed across these DRAM chips.
It multiplies the amount of data transfer between buffer chip and DRAM chips, which we refer to as the cross-chip data transfer.
The cross-chip transfer overhead, even if being overlapped with the execution of the bank PU, 
can still lead to bubbles in pipelining execution and become the bottleneck due to the limited bandwidth of rank PU.
To address these challenges, 
we first quantize the cost of cross-chip data transfer considering the configurations of LLM models, hardware parameters, and head mapping methods.
Then, we enable bubble-free pipelining with specific head mapping methods.
The analysis takes score computation of one head as an example, while context computation can follow the similar analysis.

The overheads for transferring score outputs $O_s$ can be denoted as $T_{comm}$,
following $T_{comm} = {N_{comm}}/{B_{comm}}$, where $N_{comm}$ is the amount of cross-chip data transfer, and $B_{comm}$ is the related bandwidth.
Specifically:
\begin{align*}
  {N_{comm}}={L_t}\times{N_{gqa}}\times{N_{hc}}, {B_{comm}}=({B_{rk}}\times{N_{hc}})/N_{chips}
\end{align*}
where $L_t$ is token length, $N_{gqa}$ is GQA group size, $N_{hc}$ is the number of chips allocated for the head, $B_{rk}$ is rank bandwidth, and $N_{chips}$ is total DRAM chips in the rank.
Thus, the communication time is:
\begin{align*}
  {T_{comm}}=({L_t}\times{N_{gqa}}\times{N_{chips}})/{B_{rk}}
\end{align*}
which implies that with DIMM's distributed chips, the data transfer could incur $N_{chips}\times$ additional overhead, while GQA size further exacerbates it. 

With the pipelined execution, 
the key of achieving bubble-free overlapping is $T_{comm} \le T_{comp}$, i.e., the transfer can be fully overlapped by the bank PU computation.
$T_{comp}$ can be calculated by $T_{comp} = {N_{comp}}/{B_{comp}}$, where $N_{comp}$ is the amount of bank PU data (K Cache) fetching, and $B_{comp}$ is the aggregated bank PU bandwidth.
Specifically:
\begin{align*}
  {N_{comp}}={L_t}\times{E_{h}}\times \lceil {N_{gqa}}/{N_{cmr} \rceil}, {B_{comp}}={B_{bk}}\times{N_{bk}}\times{N_{hc}}
\end{align*}
where $E_{h}$ is head embedding size, $B_{bk}$ is bank PU bandwidth, $N_{bk}$ is bank number, and $N_{cmr}$ denotes the compute-memory ratio of arithmetic units in the bank PU. 
Generally, $N_{cmr}$=1 can maximize bandwidth utilization for MHA computation, while $N_{cmr}$=8 for GQA-8 computation~\cite{park2024attacc}.
Thus, the computation time is 
\begin{align*}
  {T_{comp}}=({L_t}\times{E_{h}}\times \lceil {N_{gqa}}/{N_{cmr}} \rceil)/({B_{bk}}\times{N_{bk}}\times{N_{hc}})
\end{align*}

Considering both $T_{comm}$ and $T_{comp}$, the key for ensuring bubble-free is $T_{comm} \le T_{comp}$: 
\begin{align*}
  \frac{{L_t}\times{N_{gqa}}\times{N_{chips}}}{{B_{rk}}} \le \frac{{L_t}\times{E_{h}}\times \lceil {N_{gqa}}/{N_{cmr}} \rceil}{{B_{bk}}\times{N_{bk}}\times{N_{hc}}}
\end{align*}

To satisfy the inequation, we identify given specific model and hardware configurations, the head mapping method ($N_{hc}$) is the only tunable variable. 
This requires our head mapping to satisfy the following condition:
\begin{align}
  \label{equ:head-mapping} N_{hc} \le \frac{E_{h} \times B_{rk}\times \lceil {N_{gqa}}/{N_{cmr}} \rceil}{B_{bk} \times N_{bk} \times N_{gqa} \times N_{chips}}
\end{align}

For example, referring configurations in Table~\ref{tab:simulator} and \ref{tab:model}, our system requires ${N_{hc}}\le{8}$ for MHA ($N_{gqa} = 1$) and ${N_{hc}}\le{1}$ for GQA-8 ($N_{gqa} = 8$).
In this paper, we apply $N_{hc} = 8$ for MHA and $N_{hc} = 1$ for GQA-8, since mapping a head to more chips potentially simplifies rank-level load balance because heads of models can be mapped to more ranks.

\begin{figure}[t]
  \setlength{\abovecaptionskip}{5pt}
  \setlength{\belowcaptionskip}{-10pt}
  \centering
    \begin{minipage}[t]{\linewidth}
    \centering
    \includegraphics[width=\textwidth]{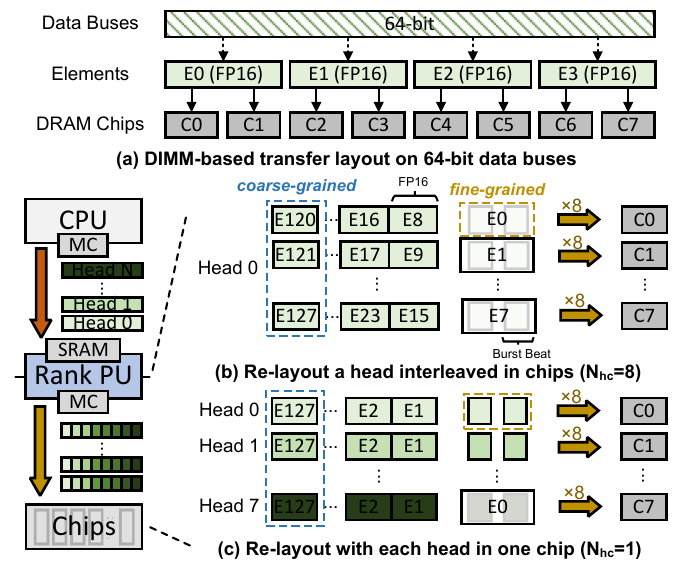}
    \footnotesize
    \end{minipage}
    \caption{\textbf{Hybrid re-layout with 8 $\times$8 chips as an example.} \textit{E0-E127: elements 0-127 in a head. C0-C7: DRAM Chip 0-7.} }
  \label{fig:design_arch_relayout}
\end{figure}

\subsection{Hybrid-grained Re-layout}
\label{sec:design-dp-relayout}

Besides specific head-mapping layouts, 
DIMMs with distributed DRAM chips also impose constraints on element layout, as shown in Fig.~\ref{fig:design_arch_relayout}-a. 
For contiguous transfers on the data buses, a single element may span multiple DRAM chips (e.g., an FP16 element across two ×8 chips), 
which fundamentally prevents PIM execution. 
Prior CPU-assisted re-layout schemes~\cite{devaux2019true,kim2021gradpim,asghari2016chameleon,lee2024pimmmu,gomez2022benchmarking} require repeated memory accesses, 
incurring non-negligible overhead (\textsection\ref{sec:eval-ablation}). 
To address this, we propose \emph{hybrid-grained relayout}, 
which leverages the rank PU’s re-layout unit to perform in-flight data transformation during communication.

The left of Fig.~\ref{fig:design_arch_relayout} shows the re-layout process with data offloading as an example. 
The QKV vectors are first buffered in the rank PU’s on-chip SRAM rather than being written directly to DRAM banks. 
After re-layout, the vectors are then stored in the DRAM chips. 
The re-layout unit addresses mismatches between element mapping and head mapping via fine-grained and coarse-grained re-layout, respectively:
First, for fine-grained re-layout, the re-layout unit ensures that each element resides on a single chip.
As shown in Fig.~\ref{fig:design_arch_relayout}, the bits of a single element are arranged in contiguous burst beats to be stored in one chip (brown dashed block).
For instance, the 16 bits of element 0 (E0) are transferred in two contiguous burst beats (within a single DDR burst), and are thus stored in Chip 0 (C0). 
Second, for coarse-grained re-layout, the re-layout unit maps each head to $N_{hc}$ chips.
Elements from a certain number of heads are scheduled together to follow the head mapping(blue dashed block).
For example, in Fig.~\ref{fig:design_arch_relayout}-b where $N_{hc}=8$, 
each burst beat contains elements from a single head after re-layout, 
so one head is transferred to 8 chips.
In Fig.~\ref{fig:design_arch_relayout}-c, elements from 8 heads are placed in one burst beat, so each head is transferred to a single chip. 
For data onloading, the re-layout performs the reverse process.

\section{Coordinated Cross-device Inference}
\label{sec:design-sys}
This section describes \sys-sys, the \sys-PIM integrated AFD system with hardware-software co-design to achieve high throughput, 
addressing the challenges of inter-device data and progress synchronization overheads.

\begin{figure}[t]
  \setlength{\abovecaptionskip}{2pt}
  \setlength{\belowcaptionskip}{-10pt}
  \centering
    \begin{minipage}[t]{\linewidth}
    \centering
    \includegraphics[width=\textwidth]{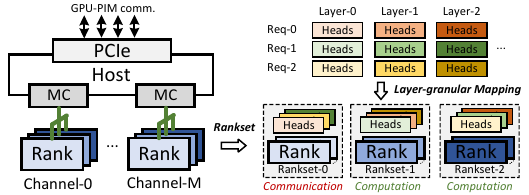}
    \footnotesize
    \end{minipage}
    \caption{\textbf{Hiding data transfer overhead with rankset-granular communication computation overlapping.}}
  \label{fig:design-zigzag-rankset}
\end{figure}

\begin{figure*}[t]
  \setlength{\abovecaptionskip}{-1pt}
  \setlength{\belowcaptionskip}{-5pt}
  \centering
    \begin{minipage}[t]{0.9\linewidth}
    \centering
    \includegraphics[width=\textwidth]{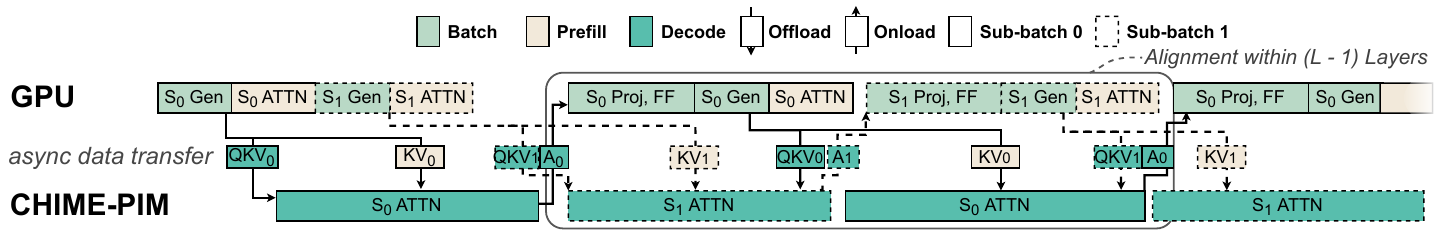}
    \footnotesize
    \end{minipage}
    \caption{\textbf{Coordinating cross-device inference with \sys-sys.} 
        \textit{\sys hides the communication cost and aligns the parallel executions across devices. ``ATTN'' denotes decoding or prefilling attention, ``Gen'' denotes QKV Generation and ``Proj, FF'' denotes the projection and feed-forward operations.}
        }
  \label{fig:design-compute-graph}
\end{figure*}

\subsection{Rankset-granular Comm-Comp Overlapping}
\label{sec:design-sys-comm}
Data synchronization between the GPU and \sys-PIM includes transferring the following data:
(1) prefilling KVs (proportional to the lengths of inputs); 
(2) decoding QKV (a token for each request), and (3) decoding attention results (a token for each request). 
Ideally, with sub-batch scheduling, the data can be transferred asynchronously, e.g., sending the decoding QKV for a sub-batch when executing the other sub-batch.
However, GPU-PIM data communication and attention computation on the \sys-PIM share the memory buses, which indicates that the communication would block the computation.
This blocking prevents asynchronous data transfer, incurring non-negligible overhead since the PCIe bandwidth~\cite{pci2025pci} could be orders of magnitude lower than that of \sys-PIM (Table~\ref{tab:simulator}).

We identify that the challenge arises from the coarse granularity, i.e., treating all ranks as a whole, for doing either communication or computation.
The key to address the challenge is finding the finest granularity of independent and concurrent communication and computation.

\myparagraph{Rankset-granular communication-computation overlapping.}
Our key observation is that, \emph{due to the shared memory buses, only one rank can be accessed in a channel at the same time during communication, while other ranks remain idle.}
This motivates our proposal of the \emph{rankset}, 
which is composed of one rank from every channel, forming the basic granularity of independent communication and computation.
In this case, a rankset is the minimum set of ranks to fully utilize all channels.
As shown in Fig.~\ref{fig:design-zigzag-rankset}, each channel has three ranks, forming three ranksets.
During the communication of one rankset, other ranksets can perform independent computation without blocking, 
which could preserve 2/3 of computational power during communication.

Moreover, we achieve \emph{rankset-granular load balance} leveraging the feature of \emph{identical KV cache sizes among layers}. 
Specifically, we store the KV cache of each request at the granularity of layers in an interleaved manner.
It ensures load balance of each rankset transfers, avoiding the rankset transferring the most data slows down the overall progress.

\subsection{Alignment Predicting Scheduling}
\label{sec:sys-scheduler}
To prevent idle bubbles caused by synchronizing the progress of parallel operations across the two devices, 
the key for \sys's scheduler is to properly select requests to form sub-batches, \emph{aligning the execution latencies of parallel operations on the two devices}.
However, LLM's auto-regressive nature makes the alignment challenging, since the execution latencies could vary with factors such as the number of processed tokens, the batch sizes, etc. 
Prior AFD systems fall short in achieving that goal:
for HBM-PIM-based AFD systems~\cite{park2024attacc,heo2024neupims}, the latencies on the PIM side could always be smaller than that on the GPU (as shown in Fig.~\ref{fig:intro-motiv}-b).
For CPU-based AFD systems~\cite{jiang2024neo}, they lack modeling the execution latencies of each operation when scheduling.
Moreover, the computation time on the CPU side can be interfered by other CPU applications and may lack predictability~\cite{christina2013paragon,liu2024jiagu,lo2015heracles}. 

\myparagraph{Opportunity of performance modeling.}
To address the challenge, \sys proposes \emph{Alignment-predicting scheduling}, 
whose key feature is modeling and predicting the execution latencies on the two devices that helps to align the parallel execution latencies. 
Leveraging the feature, it selects requests to form sub-batches, whose predicted latencies on the two devices are aligned, as shown in Fig.~\ref{fig:design-compute-graph}.
\sys exploits the following opportunities to achieve performance predictability.
First, the execution with the \sys-PIM is \emph{interference-free.}  
Second, the factors that affect the batch performances (e.g., the batch size, number of processed tokens, etc.) are known in advance.
Third, abundant prior works have explored methods for predicting the execution on the GPU~\cite{cui2021abacus,gujarati2020clockwork,yang2022infless}.

\subsection{Implementation Details for Scheduling}
\myparagraph{Model features.}
\sys describes a sub-batch $i$ with following arguments: 
$c_{p_{i}}$, the list of chunk sizes of each prefilling request;
$f_{p_{i}}$, the list of finished token numbers of each prefilling request;
$f_{d_{i}}$, the list of finished token numbers (on each rank) of each decoding request.
The latencies of both sub-batches can be modeled as follows: %
\begin{align}
  \label{equ:obj_1} &T_{{\rm GPU}_{0}} = t_{p}(c_{p_{0}}, f_{p_{0}}) + t_{\rm batch}(c_{p_{0}}, f_{d_{1}}) \\
  \label{equ:obj_2} &T_{{\rm PIM}_{0}} = t_{d}(f_{d_{0}}) + t_{\rm comm}(f_{d_{0}}, c_{p_{1}}) \\
  \label{equ:obj_3} &T_{{\rm GPU}_{1}} = t_{p}(c_{p_{1}}, f_{p_{1}}) + t_{\rm batch}(c_{p_{1}}, f_{d_{0}}) \\
  \label{equ:obj_4} &T_{{\rm PIM}_{1}} = t_{d}(f_{d_{1}}) + t_{\rm comm}(f_{d_{1}}, c_{p_{0}})
\end{align}
Equation~\ref{equ:obj_1},\ref{equ:obj_3} denotes the latency on the GPU side, which contains the latency of prefilling attention and batched FC operations.
Equation~\ref{equ:obj_2},\ref{equ:obj_4} denotes the latency on the \sys-PIM in the granularity of rank, which contains the latency of decoding attention, and overlapped data transfer overheads.

\myparagraph{Scheduling policy.}
With the following steps, the scheduling policy selects requests to form sub-batches with aligned parallel cross-device execution:
\begin{enumerate}
  \item Add one prefilling request into each sub-batch. When there is no prefilling request, skip the step.
  \item For each prefilling request added (which implies an increase in $T_{GPU}$), \sys adds $N$ decoding requests to each sub-batch and assign them to ranks in a load balancing manner. After that, it predicts $T_{PIM}$ and $T_{GPU}$ for each sub-batch. If $T_{PIM} < T_{GPU}$, it adds another $N$ decoding requests until $T_{PIM} > T_{GPU}$. 
  \item If there are remaining prefill requests, it repeats step (1) until the PIM memory is exhausted. 
  When the memory is saturated and the bubble is on the PIM side, the last prefilling request of each sub-batch is chunked, dynamically adjusting $T_{\rm GPU}$ to make it as close as possible to the $T_{\rm PIM}$ of the other sub-batch. 
\end{enumerate}

The larger the value of $N$, 
the coarser the granularity of batch execution time adjustment, 
but the more beneficial it is for achieving load balance among ranks for requests. 
In our implementation, since MHA has more heads and these heads can be evenly distributed across chips, the $N$ for MHA is set to 1, while the $N$ for GQA is set to 16.

\myparagraph{Model selection.}
\label{sec:runtime-profiling}
Now we describe how \sys models the execution latencies of various inference operations.
First, to model $T_{\rm GPU}$, we leverage Random Forest Regression (RFR) for its three advantages: capability of incremental learning, low latency, and high accuracy.
The efficiency of RFR for execution latency prediction has been proven in many prior works~\cite{liu2024jiagu,zhao2021gsight}, and it is open to use other models that feature similarly.
Second, to model $T_{\rm PIM}$, given that \sys-PIM execution is featured predictable (i.e., execution time is linearly related to the number of computed/transferred tokens),
we can use a simple and fast linear model for $t_{\rm comm}$ and each rank's $t_{d}$.
The overall $t_{d}$ is determined by the rank with the longest execution time.
The models can be described with the following formula (take sub-batch 0 as an example):
\begin{align*}
  &T_{{\rm PIM}_{0}} = {\rm Linear}(\sum{f_{d_{0}}}, {\rm len}(f_{d_{0}}), \sum{c_{p_{1}}}) \\
  &T_{{\rm GPU}_{0}} = {\rm RFR}_{p}(c_{p_{0}}, f_{p_{0}}) + {\rm RFR}_{\rm batch}(\sum{c_{p_{0}}}, {\rm len}(f_{d_{0}}))
\end{align*}

\myparagraph{Runtime profiling.}
\sys applies runtime profiling, which collects the latency information and corresponding batch information during inference.
The dataset maintained by \sys dynamically evolves with the collected data for incrementally updating the model, thereby adapting to changing execution environment.
Moreover, we evaluate the prediction models of the \sys scheduler by splitting the collected 1000 data points into training and test sets in an 8:2 ratio. 
The experimental results show that our model exhibits great performance predictability, achieving prediction relative errors of less than $\sim$1\% (median value $<$ 0.5\%).

\begin{table}[tbp] 
\setlength{\abovecaptionskip}{-1pt}
\setlength{\belowcaptionskip}{-10pt}
\caption{Simulator details.}
\begin{center}
\renewcommand\arraystretch{1.2}
\resizebox{\linewidth}{!}{
\begin{tabular}{c|c|c|c}
\toprule
\multicolumn{2}{c|}{\textbf{GPU Configuration}} & \multicolumn{2}{c}{\textbf{ACC Configuration}} \\
\hline
\textbf{Processor} & 8 A100 & \textbf{CPU} & 2TB + 406GB/s \\
\hline
\textbf{Capacity} & 640GB & \textbf{HBM-PIM} & \makecell[c]{640GB + 260.8TB/s (GPU-side) \\ \REVISE{320GB + 130.4TB/s (Extended)}}\\
\hline
\textbf{Bandwidth} & 16.3TB/s& \textbf{DIMM-PIM} & \makecell[c]{2TB + 1.6TB/s (R-PIM) \\ 2TB + 13.0TB/s (\sys)}\\
\hline
\multicolumn{4}{c}{\textbf{DIMM / DIMM-PIM: DDR4-3200}} \\
\hline
\textbf{Hierarchy} &  \multicolumn{3}{c}{ 2 Ranks (8 Chips) $\times$ 4 Bank Groups $\times$ 4 Banks}\\
\hline
\textbf{DRAM Timing} &
\multicolumn{3}{c}{\makecell[c]{BL=4:CCD=4:RRD=4/8:RCD=22:RAS=52:RP=22:RC=74:\\CL=22:WL=16:CDLR=4/12:WR=24:CCDL=8:RTP=12}} \\
\bottomrule

\multicolumn{4}{l}{ * Bank PU's compute-memory ratio $N_{cmr}$=n for GQA-n.} \\

\end{tabular}}
\label{tab:simulator}
\end{center}
\end{table}

\begin{table}[tbp]
\setlength{\abovecaptionskip}{-1pt}
\setlength{\belowcaptionskip}{-10pt}
\caption{The evaluated LLM models.}
\begin{center}
\renewcommand\arraystretch{1.2}
\resizebox{\linewidth}{!}{
\begin{tabular}{c|cccc|cc}

\toprule

\textbf{Model} & \textbf{Layers} $\mathbf{N_l}$ & \textbf{Heads} $\mathbf{N_{h}}$ & \textbf{Embedding} $\mathbf{D_e}$ & \textbf{Type} & \textbf{TP} & \textbf{DP} \\

\hline

\textbf{OPT-66B} & 64 & 72 & 9216 & MHA & 2 & 4 \\
\hline

\textbf{QWEN-72B} & 80 & 64 & 8192 & GQA-8  & 4  & 2 \\

\hline

\textbf{GPT-175B} & 96 & 96 & 12288 & MHA  & 8  & 1  \\
\bottomrule

\multicolumn{7}{l}{ * Precision: FP16; Head Embedding: $E_{h} = D_{e}/N_{h}=128$.} \\

\end{tabular}}

\label{tab:model}
\end{center}
\end{table}

\begin{table}[t]
    \setlength{\abovecaptionskip}{-1pt}
    \setlength{\belowcaptionskip}{-10pt}
    \caption{Real-world traces.}
    \begin{center}
    \renewcommand\arraystretch{1.2}
    \resizebox{\linewidth}{!}{
        \begin{tabular}{c|cccc}
            \toprule
            \textbf{Trace}           & \makecell[c]{\textbf{Avg.} $\mathbf{L_{in}}$}& \makecell[c]{\textbf{Std.} $\mathbf{L_{in}}$}  & \makecell[c]{\textbf{Avg.} $\mathbf{L_{out}}$}  & \makecell[c]{\textbf{Std.} $\mathbf{L_{out}}$} \\ 
            \hline
            \textbf{OpenR1-Math-220k}~\cite{open-r1} & 96.0              & 75.1              & 12684.1            & 8464.6             \\ 
            \hline
            \textbf{Dolphin-r1}~\cite{dophin}       & 201.9              & 563.0              & 3926.2            & 4216.0             \\ 
            \hline
            \textbf{OpenThoughts-114k-math}~\cite{OpenThoughts-114k-math} & 89.4              & 66.7              & 6366.7            & 4662.9             \\ 
            \bottomrule
            \end{tabular}
    }
    
    \label{tab:traces}
    \end{center}
    \end{table}

\section{Evaluation}
\label{sec:eval}
\begin{figure*}[htb]
  \setlength{\belowcaptionskip}{-5pt}
  \begin{minipage}[t]{0.24\linewidth}
      \centering
      \includegraphics[width=\textwidth]{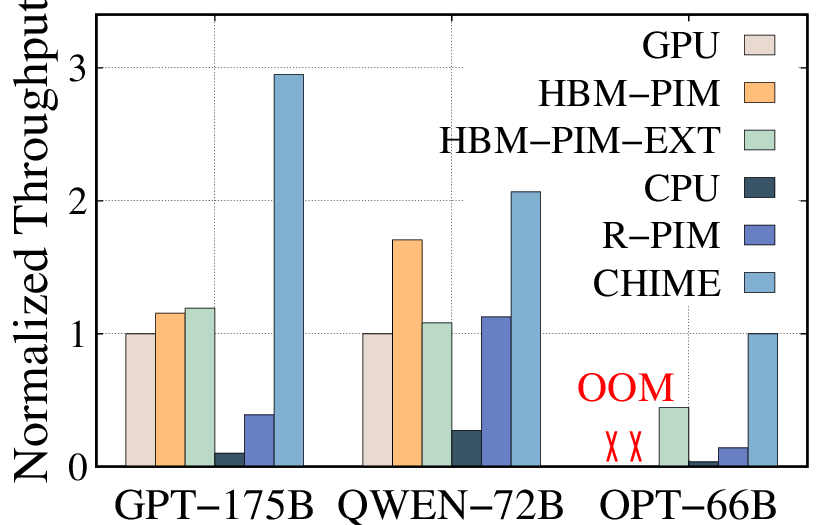}
      \footnotesize
      \textbf{(a) OpenR1.}
\end{minipage}
\begin{minipage}[t]{0.24\linewidth}
  \centering
  \includegraphics[width=\textwidth]{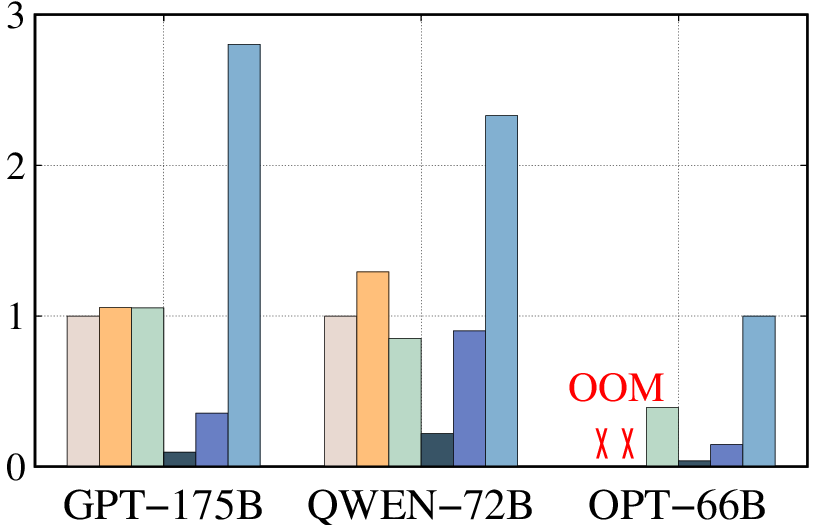}
  \footnotesize
  \textbf{(b) OpenThoughts.}
\end{minipage}
\begin{minipage}[t]{0.24\linewidth}
  \centering
  \includegraphics[width=\textwidth]{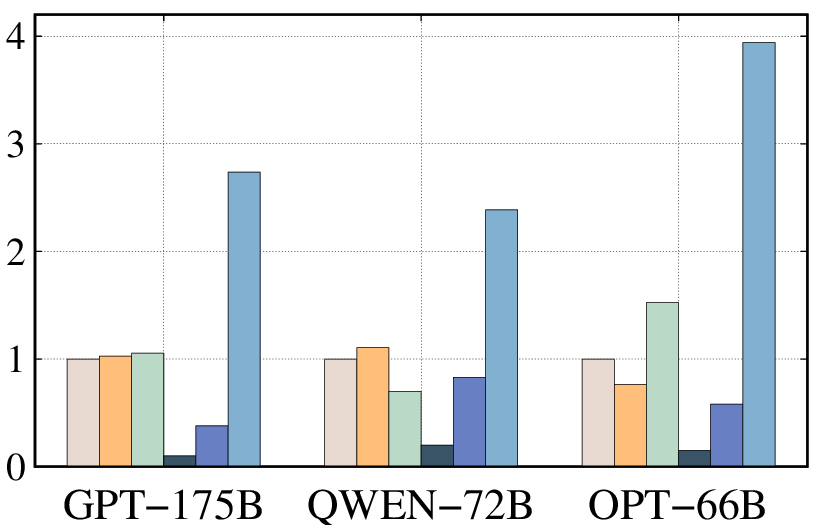}
  \footnotesize
  \textbf{(c) Dolphin.}
\end{minipage}
\begin{minipage}[t]{0.24\linewidth}
  \centering
  \includegraphics[width=\textwidth]{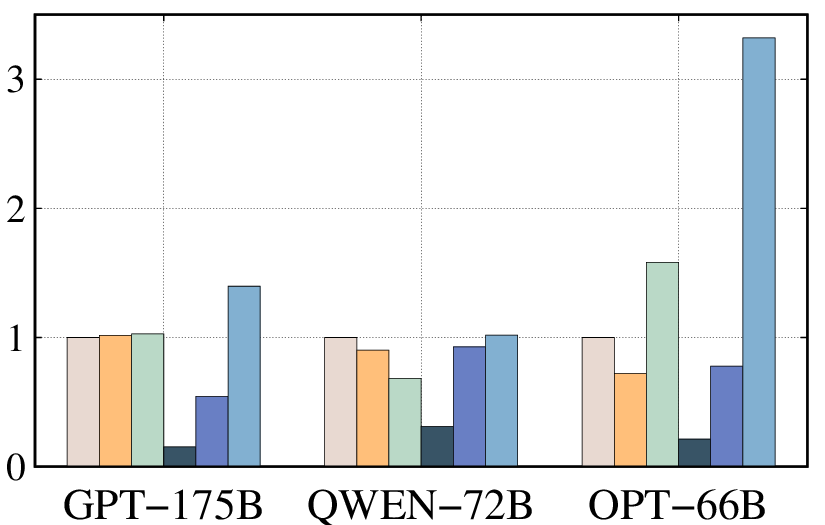}
  \footnotesize
  \textbf{\REVISE{(d) Dolphin-short.}}
\end{minipage}
\caption{\textbf{End-to-end inference throughput on real-world traces.} \normalfont{\textit{The throughput of GPU-only baseline is normalized to 1. 
For the OPT-66B model, GPU out-of-memory (OOM) errors occurred in two traces, so the throughput of \sys is normalized to 1. 
In case (d), length of requests is reduced by 10$\times$.}}}
  \label{figs:eval-end-to-end-throughput}
\end{figure*}

\subsection{Methodology}
\label{sec:eval-setup}
\myparagraph{Experimental setup.}
Following prior works~\cite{cao2025moelightening,heo2024neupims,sheng2023flexgen}, our goal is achieving \emph{high inference throughput} in the long-context and decoding-dominant scenario.
The evaluation is built upon DGX-A100~\cite{nv2023dgxa100}.
On the GPU side, 8 NVIDIA A100 GPUs, each with 5 HBM2e of 80 GB capacity, are integrated, providing a total of 156 TFLOPs on FP16.
The GPUs are connected via NVLink~\cite{wei2023nvlink}.
On the CPU side, there are 16 channels with 2 DIMMs of total 2TB capacity, equipped with the proposed PIM. 
We develop a simulator integrating
(1) AttAcc~\cite{park2024attacc}, the roofline-based simulation for GPU;
and (2) \sys-PIM-sim based on modified DRAMSim3~\cite{li2020dramsim3}, the trace-driven simulation for \sys-PIM.
\REVISE{We add PIM commands, related timing constraints, and FIFO-based scheduling methods for cycle-level simulation.}

\myparagraph{Baseline systems.}
We compare \sys with one GPU baseline and four AFD baseline systems: 
(1) \emph{GPU-only}, which executes LLM inference exclusively on GPUs. 
(2) \emph{GPU with HBM-PIM}, which equips all existing GPU HBMs with bank-level PUs.
\REVISE{(3) \emph{GPU with HBM-PIM-EXT}, which equips with extended disaggregated HBM-PIMs.
Compared to the ``HBM-PIM'' baseline, the HBM-PIMs in ``HBM-PIM-EXT'' do not store model parameters.}
(4) \emph{GPU with rank-level DIMM-PIM} (R-PIM), which equips all DIMMs with rank-level PUs.
(5) \emph{GPU with CPU offloading}, which leverages CPU as the accelerator.
Our simulations of baseline systems uniformly assume maximal accelerator-side bandwidth utilization during attention computation.
Table~\ref{tab:simulator} lists the hardware specifications. 
\REVISE{For GPU, CPU, and DIMM(-PIM) configurations, we follow the configurations in DGX-A100 systems. 
For HBM-PIM-EXT, considering that HBM2e is more than $6\times$ the price of DDR4~\cite{cost-ddr5-hbm, cost-ddr4,cost-ddr5}, we configure 20 HBM-PIM modules (320\,GB) to provide $\sim$$1/6$ of the capacity available in \sys.
}

\myparagraph{LLM models.}
We evaluate \sys on three LLM models with varying model sizes: OPT-66B, QWEN-72B, and GPT-175B, with FP16 as the data precision.
Considering both the GPU memory capacity and inter-GPU communication overhead, we apply various parallelism configurations, listed in Table~\ref{tab:model}.

\myparagraph{Workloads.}
We use three real-world LLM inference datasets: OpenR1-Math-220K~\cite{open-r1}, OpenThoughts-114k-math~\cite{OpenThoughts-114k-math}, Dolphin-r1~\cite{dophin}.
All of these datasets are used for evaluating the performance of LLM inference on long context. 
The details about the traces are shown in Table~\ref{tab:traces}.
\REVISE{
Further, to evaluate the short context scenario, we add an additional trace, 
Dolphin-short, 
which shortens each text entry in Dolphin-r1 to 1/10 of its original length.}

\subsection{End-to-end Performance}
\label{sec:eval-end-to-end}
We evaluate the end-to-end throughput for each LLM model and workload using various baselines.
For each evaluation, we randomly select and execute 1,000 requests from the traces. 
We measure the overall throughput by dividing the total number of output tokens by the total execution time.

\myparagraph{Throughput with real-world traces.}
\label{sec:eval-e2e-throughputs}
Fig.~\ref{figs:eval-end-to-end-throughput}-a,b,c presents the normalized throughput of \sys compared with \REVISE{five} baselines.
The results show that \sys achieves the highest throughput over all baselines with various settings, 
up to 5.15$\times$ higher throughput than the HBM-PIM baseline, \REVISE{3.45$\times$ than the HBM-PIM-EXT baseline}, 3.94$\times$ than the GPU-only baseline, 
and 7.21$\times$ higher than R-PIM baseline.
These improvements are attributed to the following reasons.
First, \sys achieves much larger batch sizes compared to the HBM-based baselines.
For example, for GPT-175B, \sys has 2 TB of memory on the host memory for KV cache storage, 
while GPU and HBM-PIM baselines have only about 310 GB of memory in total after deploying the model,
\REVISE{and HBM-PIM-EXT has only 320GB of memory under the same cost budget.}
Second, the CPU and R-PIM baselines suffer from the limited bandwidth.
For example, \sys offers about 8$\times$ the bandwidth of R-PIM with bank-level PUs.

We observe that HBM-PIM's sub-batch method demonstrates promising efficiency in achieving high throughput with enhanced parallelism,
though our analysis reveals certain limitations when processing Dolphin trace on OPT-66B. 
Dividing a batch into sub-batches results in smaller batch sizes, 
leading to severe GPU underutilization and performance below that of the conventional GPU-only implementation.
On the contrary, \sys benefits from much larger batch sizes and thus achieves higher throughput.

\begin{figure}[t]
  \begin{minipage}[t]{0.49\linewidth}
      \centering
      \includegraphics[width=\textwidth]{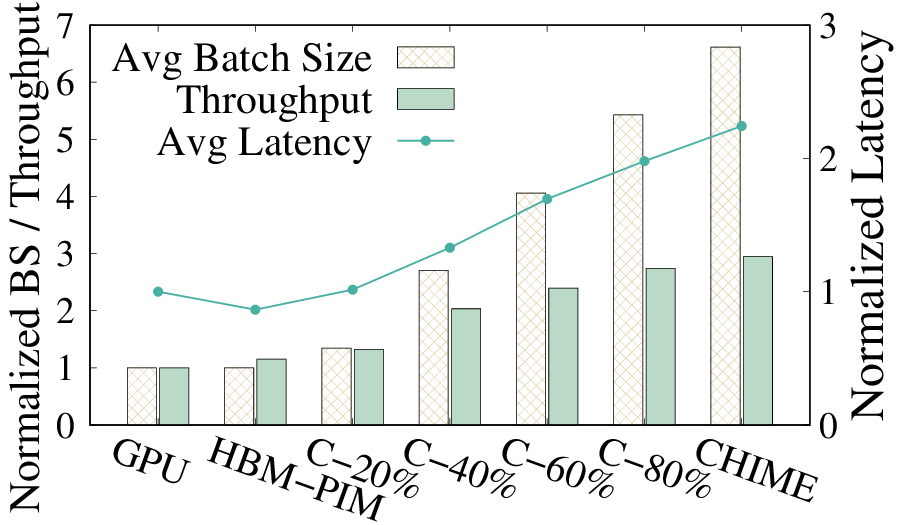}
      \footnotesize
      \textbf{(a) GPT-175B.}
  \end{minipage}
  \begin{minipage}[t]{0.49\linewidth}
      \centering
      \includegraphics[width=\textwidth]{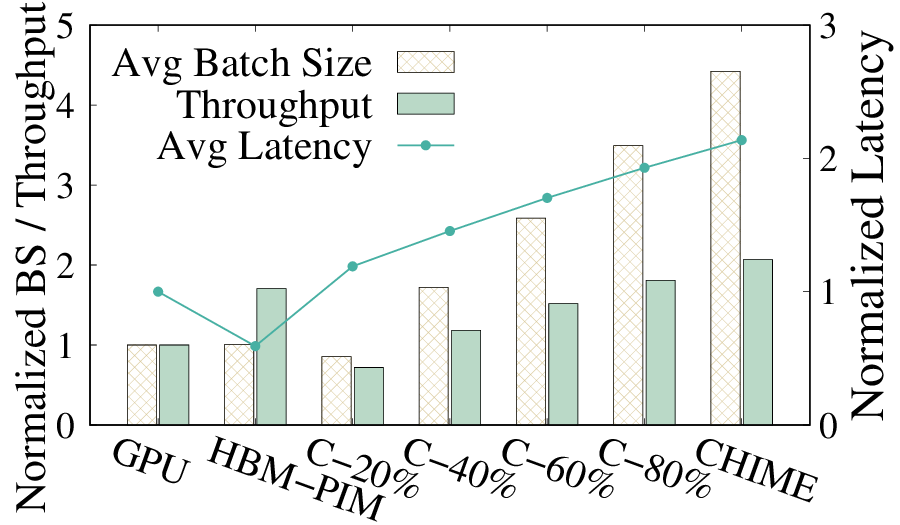}
      \footnotesize
      \textbf{(b) QWEN-72B.}
  \end{minipage}
  \caption{\textbf{\REVISE{Breakdown of performance improvement with \sys.}}
      \textit{\REVISE{The results are evaluated using the OpenR1 trace. ``C-N\%'' denotes \sys with N\% available memory capacity. ``BS'' denotes ``batch size''. ``Latency'' is ``latency per batch''. The results of GPU baseline are normalized to 1}}
  } 
  \label{figs:eval-breakdown}
\end{figure}

\myparagraph{Performance improvement breakdown.}
\REVISE{
We analyze the sources of performance improvement compared with the GPU and HBM-PIM baselines. 
Fig.\ref{figs:eval-breakdown} shows the throughput, 
average batch size (the sum of two sub-batches), 
and average latency per batch from our end-to-end evaluation. 
Compared with GPU and HBM-PIM, 
\sys increases the batch size by 6.6$\times$, 
while the latency per batch increases by only $2.2\times$. 
This indicates that the throughput improvement of \sys primarily stems from the increased batch size and the corresponding improvement in GPU utilization.
}

\REVISE{
In addition, Fig.~\ref{figs:eval-breakdown} illustrates the trend of how batch size growth affects throughput. 
We manually limit the memory capacity available to \sys and measure the resulting performance. 
As the available memory gradually increases from 10\% to 100\%, 
the batch size scales linearly. 
However, since the latency per batch also increases, the rate of throughput improvement gradually diminishes. 
This observation is consistent with the analysis in Fig.~\ref{fig:intro-motiv}, 
which shows that the marginal benefit of increasing batch size on throughput decreases.
}

\myparagraph{Performance in short context scenarios.}
\REVISE{
As shown in Fig.~\ref{figs:eval-end-to-end-throughput}-d, 
in the short context scenario, 
\sys still achieves higher throughput than both the GPU and HBM-PIM baselines by achieving higher batch sizes. 
However, 
compared to Fig.~\ref{figs:eval-end-to-end-throughput}-c, 
the advantage of \sys on each model is diminished. 
This is because in the short context scenario, 
the marginal benefit of increasing the batch size exhibits diminishing returns. 
This observation is also consistent with our analysis in \textsection\ref{sec:motiv-discussion-scenarios}.
}

\begin{figure}[tb]
  \setlength{\belowcaptionskip}{-5pt}
    \begin{minipage}[t]{0.49\linewidth}
        \centering
        \includegraphics[width=\textwidth]{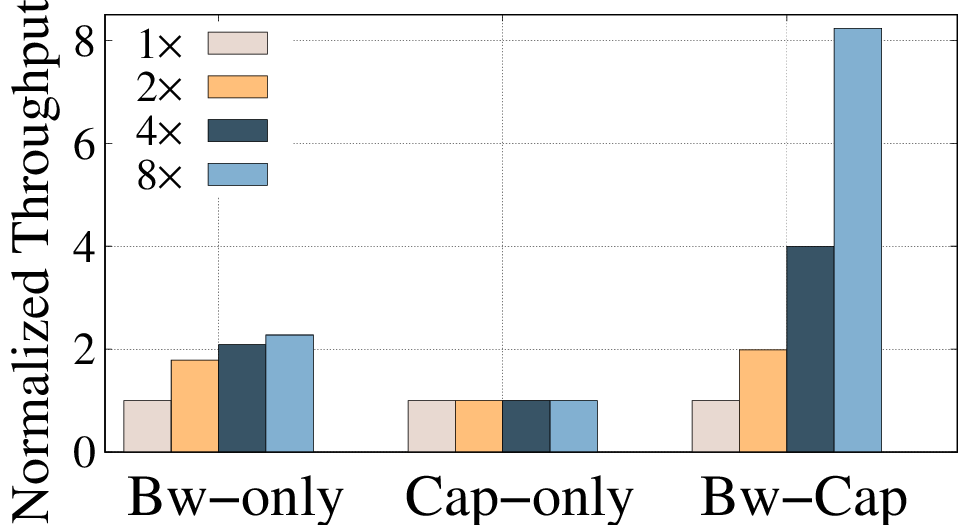}
        \footnotesize
        \textbf{(a) GPT-175B.}
    \end{minipage} 
  \begin{minipage}[t]{0.49\linewidth}
    \centering
    \includegraphics[width=\textwidth]{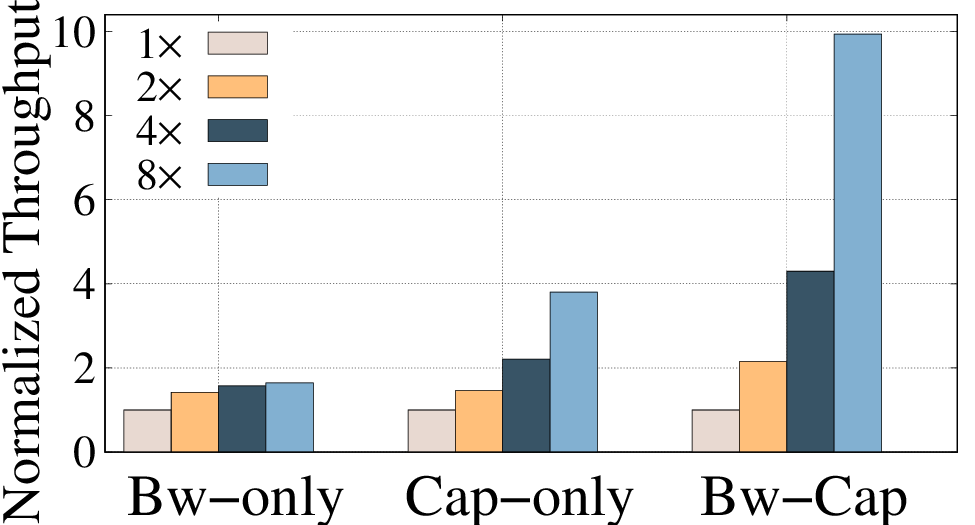}
    \footnotesize
    \textbf{(b) QWEN-72B.}
  \end{minipage} 
  \caption{\textbf{Scalability analysis.} \normalfont{\textit{The results are evaluated using the OpenR1 trace. The performance of the base memory configuration is normalized to 1.}}}
    \label{figs:eval-scalability-analysis}
  \end{figure}
  
\myparagraph{Scalability analysis.}
We further evaluate \sys performance with various \sys-PIM configurations, 
showing that performance scalability requires simultaneous scaling in memory capacity and bandwidth.
We first establish the base memory configuration as ``1 rankset, 512 GB''. 
Based on this, we scale up the memory in different ways for different baselines:
``Bw-only'' indicates that we only scale up the memory bandwidth with more ranksets but keep the capacity at 512GB, while ``Cap-only'' means we only scale up the capacity from 512GB to 4TB. 
``Bw-Cap'' implies that we scale up both the bandwidth and capacity.
We execute 1,000 requests from the trace and calculate the average throughput.

Fig.~\ref{figs:eval-scalability-analysis} shows the results of GPT-175B and QWEN-72B with OpenR1 trace. 
Results for other models and traces are similar.
It shows that expanding either bandwidth or capacity alone does not effectively improve throughput.
For example, when exclusively scaling memory capacity or bandwidth by 8$\times$ on GPT-175B, 
the throughput only increases by 2.28$\times$ and 1.01$\times$, respectively.
In contrast, when both bandwidth and capacity are enlarged, the throughput increases by 8.23$\times$. 
This demonstrates that \sys effectively leverages the value of both scalability aspects.

\begin{figure}[t]
\setlength{\abovecaptionskip}{1pt}
\setlength{\belowcaptionskip}{-5pt}
  \begin{minipage}[t]{0.49\linewidth}
      \centering
      \includegraphics[width=\textwidth]{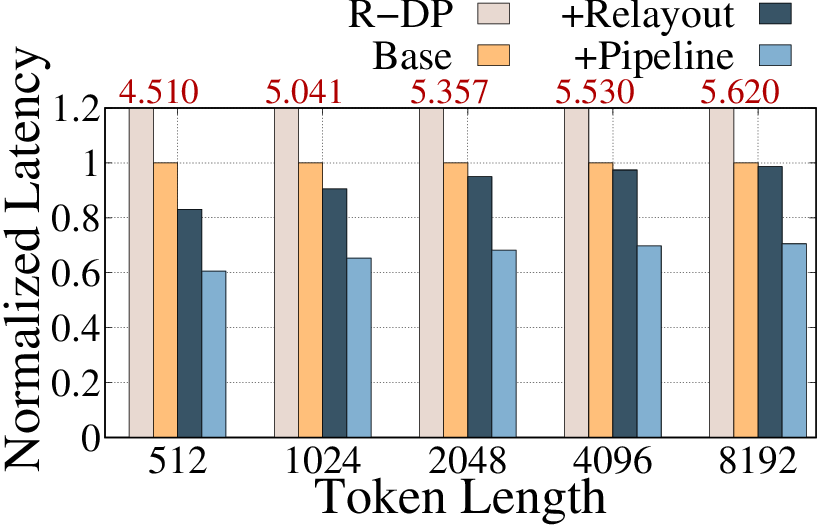}
      \footnotesize
      \textbf{(a) MHA.}
      \end{minipage}
      \begin{minipage}[t]{0.49\linewidth}
        \centering
        \includegraphics[width=\textwidth]{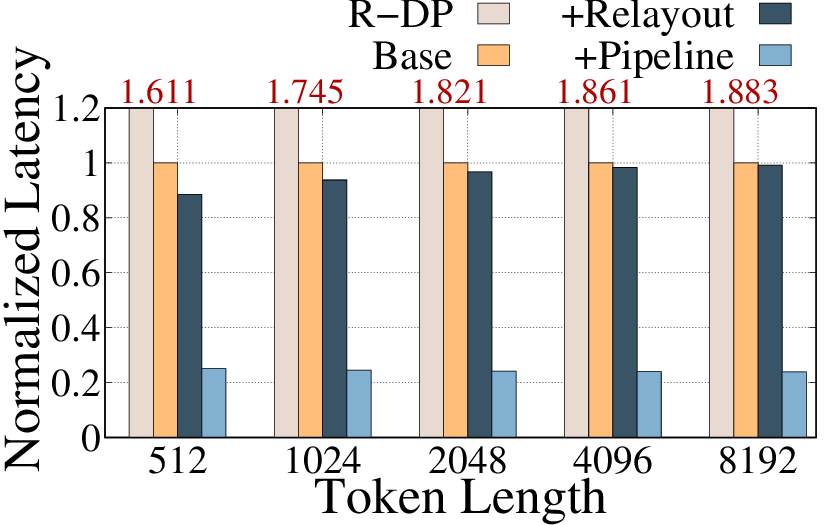}
        \footnotesize
        \textbf{(b) GQA.}
      \end{minipage}
  \caption{\textbf{Hardware ablation study.} \textit{The latency of \sys's naive implementation is normalized to 1.}} 
  \label{figs:eval-hardware-ablation}
\end{figure}

\subsection{Ablation Study}
\label{sec:eval-ablation}
\myparagraph{Ablation study for \sys-PIM.}
We analyze our hardware optimizations for: (1) bubble-free pipelining (\textsection\ref{sec:design-dp-pipeline}); (2) hybrid re-layout (\textsection\ref{sec:design-dp-relayout}). 
We implement a baseline bank-level \sys-PIM that maps head computation across chips.
Following prior work\cite{seo2025facil}, we conservatively estimate the CPU-side re-layout cost as the memory access time required to read data stored with one layout and write it into another layout.
Performance results with various token lengths are shown in Fig.~\ref{figs:eval-hardware-ablation}.
First, the bubble-free pipelining achieves about 27.9\% and 74.4\% latency reduction on MHA and GQA computation, respectively,
which is attributed to the overlapping and specific head mapping methods.
Second, the hardware hybrid re-layout enables up to 17\% latency reduction.
The reason why the proportion of gain decreases as token length increases is that the attention computation gradually dominates the overall execution time.
Nonetheless, we identify that it is still needed since the re-layout overheads accumulate in each layer of each token generation.
With all hardware optimizations, \syscolor shows 1.42--4.18$\times$ speedup. 
In addition, all bank-level implementations achieve over 1.5$\times$ speedup than the rank-level DIMM-PIM (R-DP).

\begin{figure}[t]
  \setlength{\abovecaptionskip}{0.5pt}
  \setlength{\belowcaptionskip}{-5pt}
  \begin{minipage}[t]{0.49\linewidth}
      \centering
      \includegraphics[width=\textwidth]{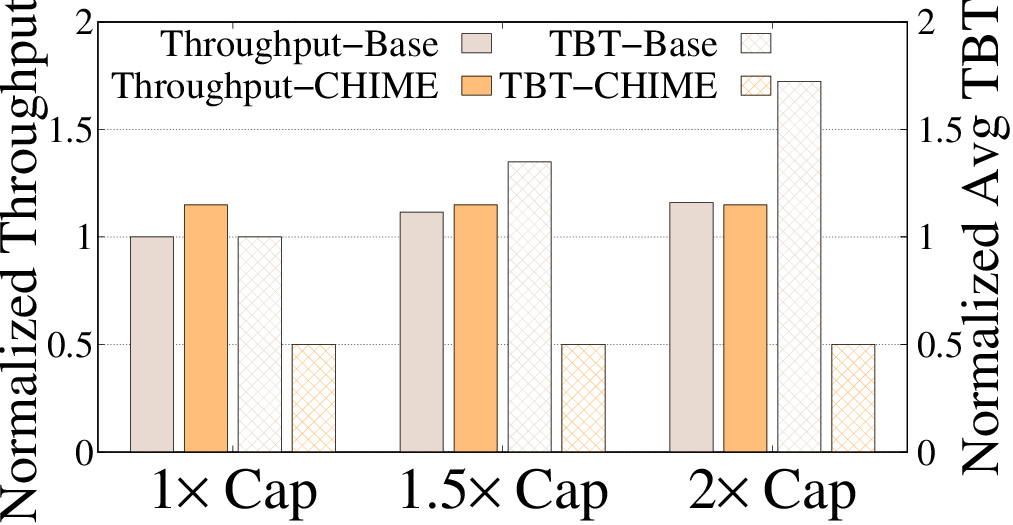}
      \footnotesize
      \textbf{(a) GPT-175B.}
      \end{minipage}
      \begin{minipage}[t]{0.49\linewidth}
        \centering
        \includegraphics[width=\textwidth]{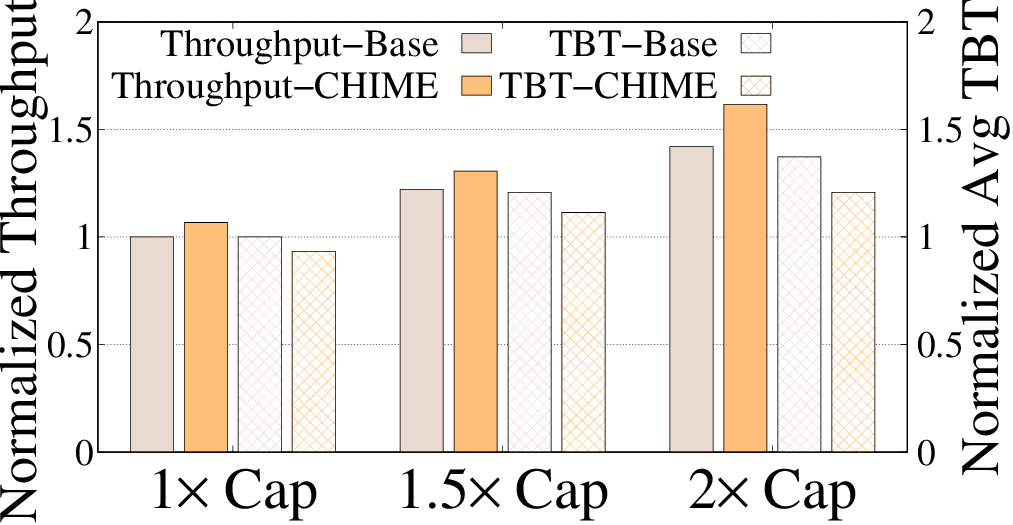}
        \footnotesize
        \textbf{(b) QWEN-72B.}
      \end{minipage}
  \caption{\textbf{Improved resource efficiency with \sys-sys.} 
      \textit{``1$\times$ Cap'' denotes the basic memory capacity (2TB for GPT-175B and 1TB for QWEN-72B) according to Table~\ref{tab:model}. 
      The results are evaluated using the OpenR1 trace.}
  } 
  \label{figs:eval-scheduler}
\end{figure}

\myparagraph{Ablation study for scheduler.}
We evaluate how \sys's scheduler effectively improves the resource utilization.
Fig.~\ref{figs:eval-scheduler} shows the average throughput and Time-between-Tokens (TBTs) of different scheduling methods on the left and right y-axes, respectively. 
The baseline represents the scheduling policy that prioritizes filling the capacity of \sys-PIM, 
while \sys denotes the alignment-predicting scheduling. 
The results show that \sys's scheduler can significantly reduce latency by up to 70.93\% without sacrificing the throughput, 
showing its ability of eliminating idle bubbles on the GPU side.
\sys even slightly improves the throughput by aligning the batches with prefilling requests and achieving better load balancing.
Moreover, for the MHA model, if we increase the capacity of \sys-PIM, 
the baseline selects larger batch sizes and causes higher TBTs without improving the throughput (since attention becomes the bottleneck), 
while \sys can avoid generating bubbles and prevent the growth of TBT.
\sys performs better with MHA models, because with the GQA model, achieving the attention bottleneck requires selecting requests that occupy larger memory capacity, or it may not even achieve the attention bottleneck.

\begin{figure}[t]
  \setlength{\abovecaptionskip}{1pt}
  \setlength{\belowcaptionskip}{-5pt}
  \centering
    \begin{minipage}[t]{0.68\linewidth}
        \centering
        \includegraphics[width=\textwidth]{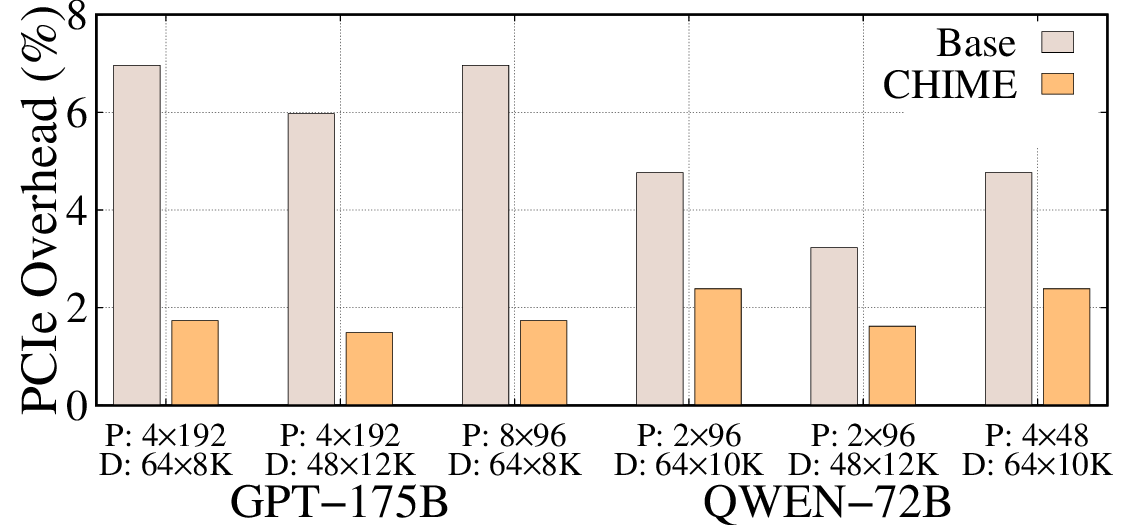}
        \textbf{(a) PCIe overhead.}
        \footnotesize
    \end{minipage}
    \begin{minipage}[t]{0.30\linewidth}
        \centering
        \includegraphics[width=\textwidth]{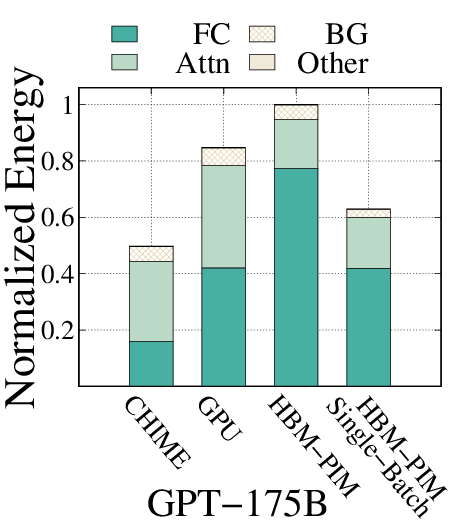}
        \textbf{\REVISE{(b) Energy.}}
        \footnotesize
    \end{minipage}
  \caption{\textbf{PCIe overhead and energy consumption.}
        \textit{``P'' and ``D'' in (a) denotes the batch sizes and token lengths of prefilling and decoding requests in a batch. The energy consumption of ``GPU'' in (b) is normalized to 1.}
     }
  \label{figs:eval-pcie-energy}
\end{figure}

\myparagraph{PCIe overhead analysis.}
We evaluate how the cross-device PCIe communication degrades the throughput with or without rankset-granular communication computation overlapping.
As shown in Fig.~\ref{figs:eval-pcie-energy}, applying the optimization could reduce the overhead by up to 75.08\% with different batch configurations.
This indicates that with the 4 ranksets of DGX-A100, most PCIe overheads are hidden with independent communication and computation at the rankset granularity.

\subsection{Cost Analysis}
\myparagraph{Energy consumption.}
\REVISE{
We evaluate the energy consumption of \sys with GPU and HBM-PIM.
For GPU-side energy, we follow the evaluation methods in AttAcc simulators. 
For PIM-side energy, 
first, we use power parameters (e.g., VDD, IDD) and timing parameters (e.g., tRAS, tRP) to calculate energy of ACT, RD, WR, and so on~\cite{micronddr4power1,micronddr4power2}. 
Second, we follow the scaling methods~\cite{mike2017fgdram,park2021trim,park2024attacc} to calculate read/write energy (e.g., MAC, Load shared buffer) inside the DRAM chips (or DRAM die); 
Third, we obtain the chip I/O energy from CACTI-IO~\cite{Jouppi2012cactiio}.
We provide three baselines: GPU, HBM-PIM, and HBM-PIM without sub-batch scheduling.}

\REVISE{
Fig.~\ref{figs:eval-pcie-energy}-b details the energy breakdown for GPT-175B on openR1 traces, 
where FC and attention computations dominate. 
First, \sys achieves a ~40\% total energy reduction, 
primarily by executing FC on the GPU with larger batch sizes, which reduces the number of weight loading. 
Second, HBM-PIM with sub-batch scheduling consumes more energy, since it halves the batch size.
Third, for attention computation, \sys-PIM's multi-chip distributed computation consumes energy comparable to the GPU but higher than HBM-PIM. 
}

\myparagraph{Hardware overhead.}
\REVISE{
We evaluate the area overhead of bank PUs, whose in-DRAM-chip integration directly impacts memory capacity.
Assuming command decoding and data paths reuse existing DRAM logic, we synthesize the design using a TSMC 28nm process. 
A single bank PU and the shared buffer consume 0.0032 $mm^2$ and 0.051 $mm^2$, respectively. 
Accounting for density differences between logic and DRAM processes, we estimate that implementation in a 1z-nm DRAM node would roughly double the synthesized area. 
While efficiently supporting GQA requires scaling the bank PU proportionally, we consider this overhead highly acceptable for two reasons.
First, the massive capacity of DIMMs, effectively amortizes the area cost. 
Further, emerging 3D integration technologies can further mitigate or eliminate this die area penalty.}

\begin{figure}[t]
  \setlength{\abovecaptionskip}{2pt}
  \setlength{\belowcaptionskip}{-10pt}
  \centering
    \begin{minipage}[t]{\linewidth}
    \centering
    \includegraphics[width=\textwidth]{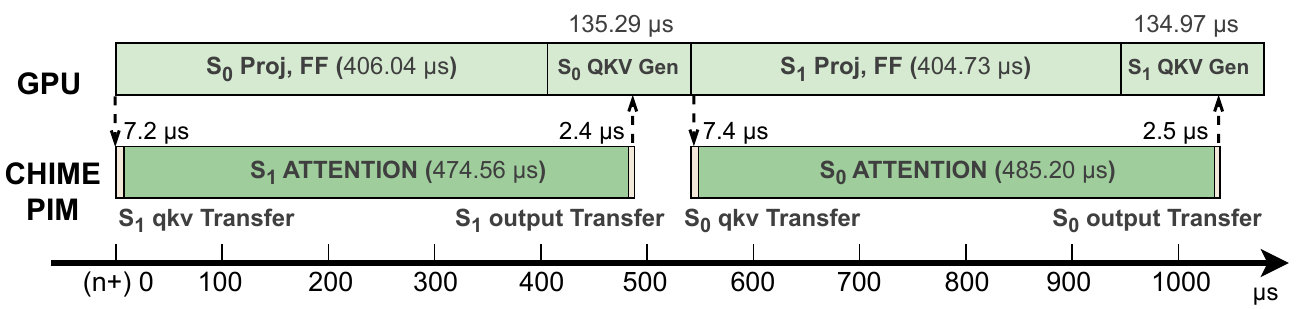}
    \footnotesize
    \end{minipage}
    \caption{\textbf{\REVISE{Decode execution timeline of CHIME pipeline for single LLM layer.}}
    \textit{\REVISE{The batch is executed with GPT-175B and the OpenR1 trace.}}}
  \label{fig:eval-pipeline-timeline}
\end{figure}

\myparagraph{Timeline breakdown analysis.}
\REVISE{
As shown in Fig.~\ref{fig:eval-pipeline-timeline}, 
we extract a representative batch from the actual execution of the end-to-end evaluation as an example and perform a detailed timeline breakdown analysis. 
We observe that with our designs, the computation on the GPU and PIM sides is effectively pipelined with minor bubbles.
}

\section{Related Work}

\label{subs:motiv-related-work}

\myparagraph{PIM/PNM architecture for LLM inference.}
In addition to PIM designs~\cite{park2024attacc,heo2024neupims} leveraged in AFD systems for batched inference,
IANUS~\cite{seo2024ianus,li2025h2llm} accelerates non-batch inference with distinct mapping methods.
For long context scenarios, LoL-PIM~\cite{kwon2025lolpim}, CXL-PNM~\cite{park2024cxlpnm} and HPU~\cite{rhee2025hpu} propose scalable capacity and bandwidth,
in which DRM can guide their configurations to achieve optimal performance.
Duplex~\cite{yun2024duplex} targets Mixture of Experts (MoE) inference. 
The DRM can be readily extended to analyze MoE models, as their FC layers exhibit performance patterns similar to those of dense models.
\REVISE{
Papi~\cite{he2025papi} proposes dynamic scheduling methods on various FC devices under different batch sizes.
We identify that it is orthogonal to our work, and these devices represent different Line-FCs in the DRM model,
which show high performance potential under several real-world scenarios, such as low request per second.
CENT~\cite{gu2025cent} proposes scalable PIM-only inference and could achieve high performance in various scenarios.
In AFD inference scenarios, the cost efficiency of its GDDR6-PIM degrades, as shown in Fig.~\ref{fig:motiv-memory-bottleneck}.
}

\myparagraph{Systems with enhanced capacity for KV cache.}
FastDecode~\cite{he2024fastdecode} and NEO~\cite{jiang2024neo} offload the KV cache to the host memory and compute attention with the CPU. 
InstAttention~\cite{pan2025instattention} offloads attention with computational storage drives, utilizing the internal bandwidth of the SSD.
These systems suffer limited bandwidth for attention computation, as analyzed in \textsection\ref{s:motiv-analytic-model}.
Other works such as CachedAttention~\cite{Gao2024cachedattention} and FlexGen~\cite{sheng2023flexgen} utilizes host memory to store the KV cache, and fetch it back to GPU for attention computation.
Their latencies is limited by the PCIe bandwidth.

\myparagraph{GPU-only AFD systems.}
Some prior works disaggregate Attention and FC on heterogeneous GPUs~\cite{step3blog,step3system,zhu2025megascaleinfer}.
Our DRM-based methodology can help them more effectively balance compute between Attention and FC. 
While GPUs offer superior generality and software compatibility, DIMM-PIM provides substantial cost advantages.

\section{Conclusion}
\label{s:conclusion}

This paper 
presents
the first general AFD performance model, Disaggregated Roofline Model, from which we conclude the ``Liebig's Law'' that guides the design of \sys,
a hardware-software co-designed AFD system with DIMM-PIM that offers scalable memory capacity and bandwidth for LLM inference.
\sys can enable efficient DIMM-PIM attention computation and maximizes the resource utilization for cross-device inference.
Evaluations show \sys achieves significantly higher throughput, establishing a new paradigm for efficient long-context LLM inference.

\section*{Acknowledgments}  
We thank the anonymous ISCA reviewers for their constructive feedback.
This work was supported in part by the National Natural Science Foundation of China (No. 62302300, 62432010, 62472279, and U24B20165), 
the Fundamental and Interdisciplinary Disciplines Breakthrough Plan of the Ministry of Education of China (JYB2025XDXM122),
the Fundamental Research Funds for the Central Universities,
and the Advanced Research Center for Agent-Oriented OS (TC20260106010).
Corresponding authors: Dong Du (\url{dd\_nirvana@sjtu.edu.cn}) and Naifeng Jing (\url{sjtuj@sjtu.edu.cn}).

\bibliographystyle{IEEEtranS}
\bibliography{refs}

\end{document}